\documentclass[twocolumn,epsfig,a4paper,amsmath,amssymb,showpacs,prl,superscriptaddress]{revtex4-1}
\usepackage{multirow}
\usepackage{amssymb}
\usepackage{amsmath}
\usepackage{amsfonts}
\usepackage{latexsym}
\usepackage{graphicx}
\usepackage{caption}
\usepackage{subcaption}
\usepackage{dcolumn}
\usepackage{bm}
\usepackage{ulem}
\usepackage{color}
\usepackage{soul}
\usepackage{physics}
\usepackage{bbold}
\usepackage{gensymb}

\graphicspath{ {./figs/} }

\newcommand{\be}{\begin{equation}}
\newcommand{\ee}{\end{equation}}
\newcommand{\ba}{\begin{eqnarray}}
\newcommand{\ea}{\end{eqnarray}}
\newcommand{\baa}{\begin{eqnarray*}}
\newcommand{\eaa}{\end{eqnarray*}}
\newcommand{\Lim}[1]{\raisebox{0.5ex}{\scalebox{0.8}{$\displaystyle \lim_{#1}\;$}}}

\newcommand{\dg}{^{\dagger}}
\newcommand{\mI}{{\mathbb 1}}
\newcommand{\me}{\mathrm{e}}
\newcommand{\bl}{\left (}
\newcommand{\br}{\right )}
\newcommand{\Vbar}{\overline{\overline{V}}}
\newcommand{\Gbar}{\overline{\overline{G}}}
\newcommand{\Tbar}{\overline{\overline{T}}}
\newcommand{\gbar}{\overline{\overline{\gamma}}}

\newcommand{\Gammabar}{\overline{\overline{\Gamma}}}

\begin{document}

\title{Spin Transport and Accumulation in a 2D Weyl Fermion System}
\author{T.~Tzen Ong}
\affiliation{RIKEN Center for Emergent Matter Science (CEMS), Saitama 351-0198, Japan}
\affiliation{Department of Applied Physics, University of Tokyo, Tokyo 113-8656, Japan}
\author{Naoto Nagaosa}
\affiliation{RIKEN Center for Emergent Matter Science (CEMS), Saitama 351-0198, Japan}
\affiliation{Department of Applied Physics, University of Tokyo, Tokyo 113-8656, Japan}
\date{\today}

\begin{abstract}
In this work, we study the spin Hall effect and Rashba-Edelstein effect of a 2D Weyl fermion system in the clean limit using the Kubo formalism. Spin transport is solely due to the spin-torque current in this strongly spin-orbit coupled (SOC) system, and chiral spin-flip scattering off non-SOC scalar impurities, with potential strength $V$ and size $a$,  gives rise to a skew-scattering mechanism for the spin Hall effect. The key result is that the resultant spin-Hall angle has a fixed sign, with $\theta^{SH} \sim O \left(\tfrac{V^2}{v_F^2/a^2} (k_F a)^4 \right)$ being a strongly-dependent function of $k_F a$, with $k_F$ and $v_F$ being the Fermi wave-vector and Fermi velocity respectively. This, therefore, allows for the possibility of tuning the SHE by adjusting the Fermi energy or impurity size. 
\end{abstract}

\maketitle

The spin Hall effect (SHE) has a long and rich history, starting with the initial proposal of asymmetric Mott scattering by Dyakonov and Perel \cite{DyakonovJETP1971,DyakonovPhysLettA1971}. This extrinsic mechanism was re-introduced in 1999\cite{HirschPRL1999,ZhangPRL2000}, while an intrinsic SHE was first proposed in 2003\cite{MurakamiNagaosaZhangScience2003,SinovaMacDonaldPRL2004}. The proposal of a two-dimensional (2D) $Z_2$-protected Quantum Spin Hall (QSH) state\cite{KaneMelePRL2005}, and its successful prediction in HgTe/CdTe quantum well \cite{BernevigZhangScience2006} quickly followed; thus giving rise to a new field of topological materials\cite{KaneHasanRMP2010, ZhangQiRMP2011}, which now include 2D QSH states~\cite{[{For a review of SHE and QSHE, see }]NagaosaMurakamiCSST2011}, 3D topological insulators (TI)\cite{KaneFuPRL2007, MooreBalentsPRB2007}, topological Kondo insulators\cite{DzeroColemanPRL2010,XuDingNatComm2014} and Weyl semi-metals\cite{VishwanathWanPRB2011}. 

One of the most striking characteristic of 3D TI materials is the existence of spin-momentum locked chiral Weyl fermions on the surfaces, which are expected to provide highly efficient spin-charge conversion\cite{FertNatComms2013, ShiomiSaitohPRL2014}, via the spin Hall effect or spin accumulation in the Rashba-Edelsten effect\cite{EdelsteinSSC1990}. Hence, there is a strong interest in spintronic TI heterostructures, with many theoretical works\cite{ShankarMondalPRB2010, DasSarmaCulcerPRB2010,BurkovHawthornPRL2010,FranzGaratePRL2010,NagaosaZangPRB2010,NagaosaBranislavPRL2012}, discussing a plethora of spin-charge phenomena, including magnetoresistance effects, inverse spin-galvanic effect, and spin-transfer torque, which have stimulated a flurry of experimental efforts\cite{RalphMellnikNature2014, LiJonkerNNano2014, ShiomiSaitohPRL2014, AndoNanoLett2014, KangNatMat2014, TokuraKondouNatPhys2016}.

In heavy-metal/ ferromagnet systems, e.g. FePt/Au, a giant spin Hall angle (SHA) of $\sim 0.1$ has been reported\cite{SekiTakanashiNatMat2008}, which has been interpreted as resonant skew-scattering off the Fe impurities\cite{GuNagaosaPRL2009}. However, recent experiments on TI heterostructures\cite{KangNatMat2014,RalphMellnikNature2014} have reported values of $\tan \theta^{SH} > 100$\%, with combined surface and bulk contributions. In order to disentangle the surface Weyl fermion contribution from the bulk bands, a Cu-layer inserted TI/Cu/ferromagnet heterostructure has recently been engineered, with $\tan \theta^{SH} \sim 50$\% \cite{TokuraKondouNatPhys2016}. 

Similar to the anomalous Hall effect, there are both intrinsic Berry curvature and extrinsic scattering contributions to the SHE. For systems with weak spin-orbit coupling (SOC), it has been shown\cite{HalperinRashbaPRL2005} that the extrinsic skew scattering mechanism dominates in the clean limit; hence, the spin Hall conductivity $\sigma^z_{xy}$ scales with the longitudinal conductivity $\sigma_{yy}$, and the SHA, $\theta^{SH} = \tfrac{\sigma^z_{xy}}{\sigma_{yy}}$ is a well-defined measure of the SHE. The Rasha-Edelstein effect is a closely related transport-driven spin accumulation phenomena, which also scales with $\sigma_{yy}$ in the clean limit; the spin accumulation $\langle S^i \rangle = \sigma^i_{\alpha} E_{\alpha}$ is proportional to the applied electric field $E_{\alpha}$ (along $\alpha$-direction) with a coefficient $\sigma^i_{\alpha}$. For the strongly SOC-coupled Weyl system considered here, the main results are that due to spin-momentum locking, chiral spin-flip scattering off non-magnetic impurities drives an $O(\tfrac{1}{n_i})$ skew-scattering mechanism, and that Rashba-Edelstein is an $O(\tfrac{1}{\gamma_t})$ effect; here, $n_i$ is the impurity concentration and $\gamma_t$ is the transport scattering rate.

We adopt the Kubo formula framework for calculating $\sigma_{yy}$, $\sigma^{z}_{xy}$ and $\sigma^{i}_{y}$, given by the retarded current-current correlation functions, $\sigma_{yy} = - \Lim{\omega \rightarrow 0} \Lim{\vec{k} \rightarrow 0} Im \left[ \frac{\pi_{yy}(\vec{k}, \omega)}{\omega} \right]$, $\sigma^{z}_{xy} = -\Lim{\omega \rightarrow 0} \Lim{\vec{k} \rightarrow 0} Im \left[ \frac{\pi^{z}_{xy}(\vec{k}, \omega)}{\omega} \right]$, and $\sigma^{i}_{y} = -\Lim{\omega \rightarrow 0} \Lim{\vec{k} \rightarrow 0} Im \left[ \frac{\pi^{i}_{y}(\vec{k}, \omega)}{\omega} \right]$; where, $\pi_{yy}(\vec{k}, \omega)$, $\pi^{z}_{xy}(\vec{k}, \omega)$, and $\pi^{i}_{y}(\vec{k}, \omega)$ are the current-current, spin current-current and spin accumulation-current correlation functions respectively.

In spin-orbit coupled systems, the proper definition of the spin current is more subtle as spin is not a conserved quantity. Ref.~\cite{NiuPRL2006} presented a bulk conserved spin current that satisfies a continuity equation, $\tfrac{d S^z}{d t} + \nabla \cdot (\vec{J}_s + \vec{P}_{\tau}) = 0 $, with an additional spin-torque density term, $\nabla \cdot \vec{P}_{\tau}^i = \tfrac{i}{\hbar} [S^i, H^0]$, as well as the conventional spin current $\vec{j}^{z}_s = \psi\dg \tfrac{1}{2} \{\vec{v}, S^z\} \psi$. Hence, the transport spin current is the sum of a spin-polarized and a spin-torque current, $\vec{\mathcal{J}}^i_s = \vec{j}^i_s + \vec{P}^i_{\tau}$, succintly expressed as the time-derivative of a spin-dipole operator, $\hat{\mathcal{J}}_s = \tfrac{d (\hat{\vec{r}} \hat{\vec{S}}) }{d t}$. As pointed out by several groups\cite{MolenkampPRB2003, HalperinMischenkoPRL2004, NagaosaSugimotoPRB2006}, there is no finite conventional spin current for Weyl systems; hence, spin transport for Weyl fermions is solely due to the spin-torque density $P_{\tau}$ coming from quantum-mechanical evolution of the electron spin.  
\begin{figure}
\begin{center}
\begin{subfigure}{0.5 \columnwidth}
\makebox[10pt][l]{(a)}
\includegraphics[width=\columnwidth]{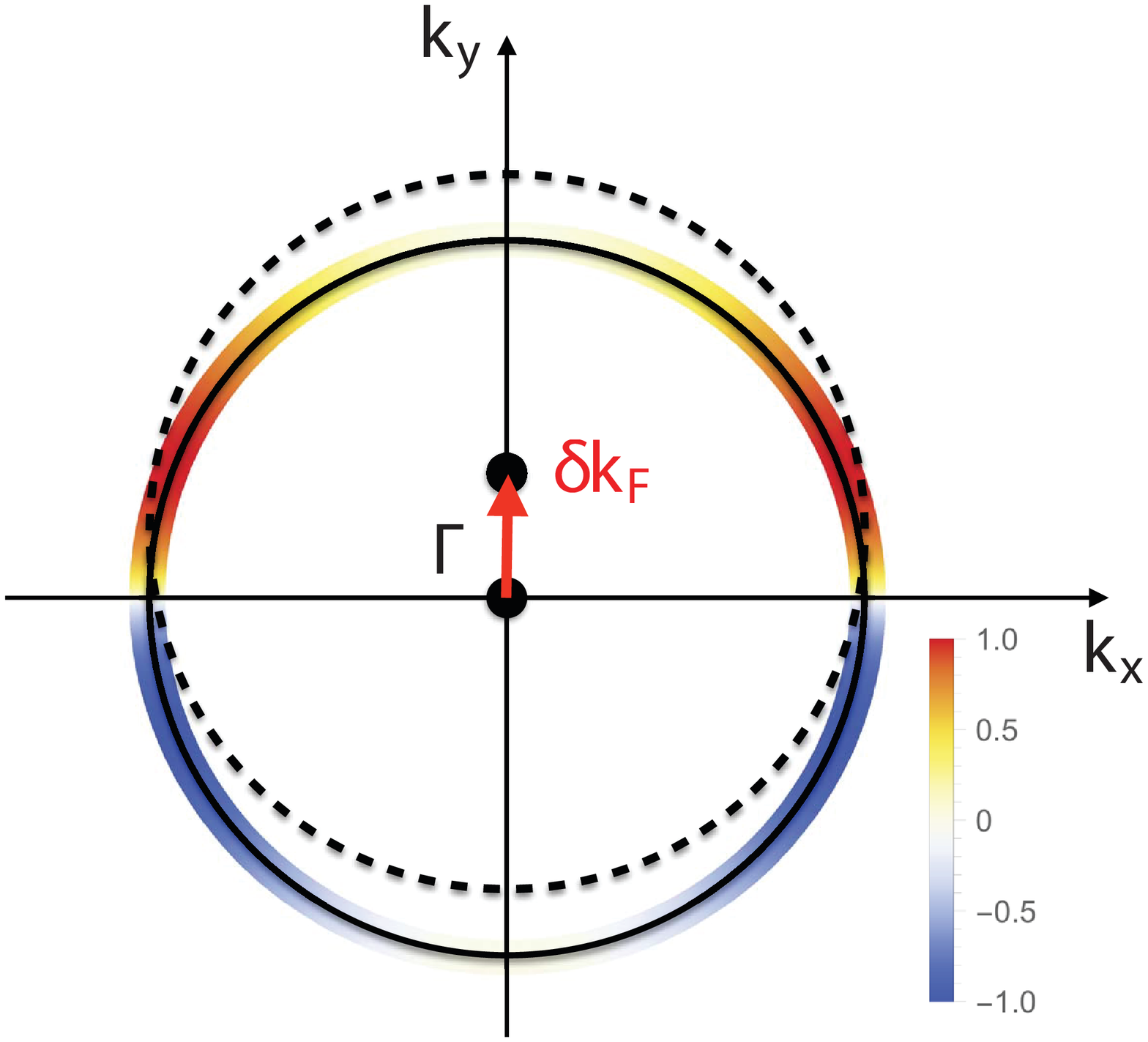}
\end{subfigure}%
\begin{subfigure}{0.45 \columnwidth}
\makebox[10pt][l]{(b)}
\includegraphics[width=\columnwidth]{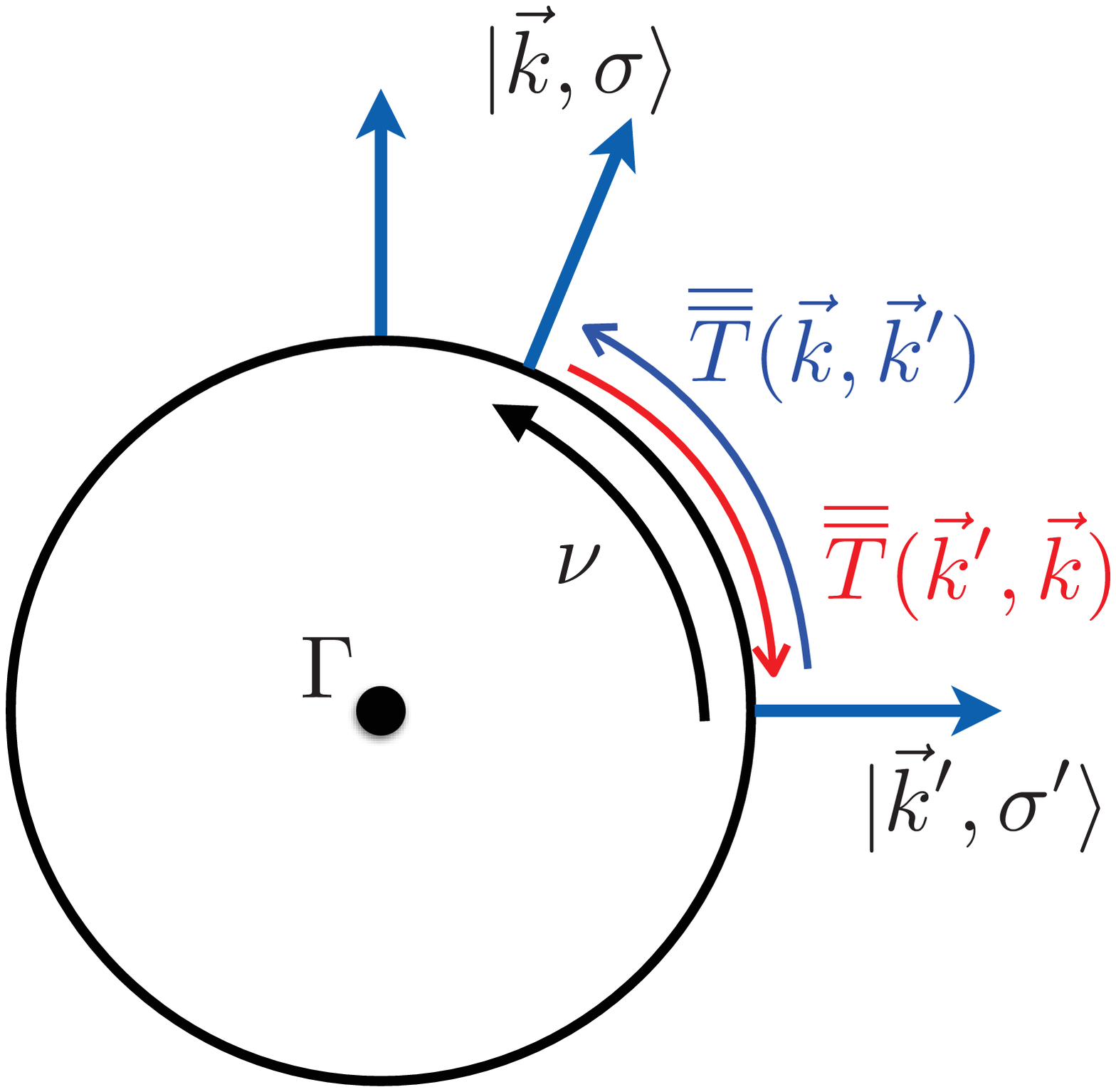}
\end{subfigure}
\end{center}
\captionsetup{justification=RaggedRight, singlelinecheck=false}
\caption{Fig.~(a) shows a colour density plot of the FS contribution to the Rashba-Edelstein effect $\langle \sigma^{y}_y \rangle$ (Eq.~\ref{eqn: sigmaiy}). When the FS is shifted by $\delta k_y = e E_y \tau_t$ due to an external electric field $E_{y}$, the non-equilibrium distribution gives rise to a net $\langle S^y \rangle$.  Fig.~(b) illustrates spin-dependent skew scattering, $\Tbar_{\sigma, \sigma'}(\vec{k}, \vec{k}')$ and $\Tbar_{\sigma', \sigma}(\vec{k}', \vec{k})$ having positive ($\nu$) and negative ($-\nu$) chirality respectively, with the helical Weyl fermions defining positive ($\nu$) chirality.}
\label{fig: Fermi surface}
\end{figure}
 
\begin{figure}[bht!]
\begin{center}
\includegraphics[trim=0mm 0mm 0mm 0mm, clip, width=\columnwidth]{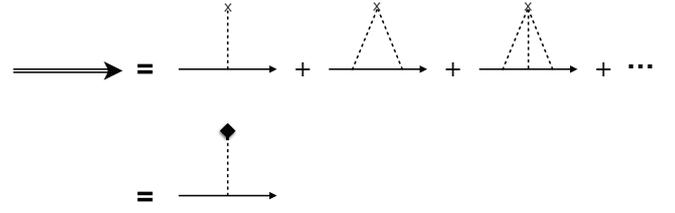}
\end{center}
\captionsetup{justification=RaggedRight, singlelinecheck=false}
\caption{\label{fig: Geff Feyn diag} 
Feynman diagram for  $\Gbar_{eff}(\vec{k}, \vec{k}', \sigma, \sigma')$ that sums up the infinite set of scattering events from a single impurity. This is captured by the $\Tbar$-matrix, which is represented by the diamond symbol in the second line above.}
\end{figure}

We consider elastic scattering near the Fermi energy, $E_F$, of 2D Weyl fermions (Dresselhaus-type $v_F \, \vec{k} \cdot \vec{\sigma}$ system) from a dilute ($n_i \ll 1$) random distribution of non-magnetic impurities, with scattering off each impurity given by $H^{imp} = \sum_{\vec{r}} c\dg_{\sigma}(\vec{r}) V \mathrm{e}^{-\frac{|\vec{r}|^2}{a^2}} c_{\sigma}(\vec{r})$, with impurity size $a$. Note that the results can be easily translated into the Rashba-type $v_F \, \hat{z} \times \vec{k} \cdot \vec{\sigma}$ case via rotation of the momentum by $90^{\degree}$. Choosing the chemical potential $\mu$ to lie in the upper helical band, we obtain the following Hamiltonian as,
\ba
\label{eqn: Hamiltonian}
H & = & H^0 + H^{imp} \\
H^0 & = & \sum_{\vec{k}, \alpha, \beta} c\dg_{\vec{k}, \alpha}  v_F  \vec{k} \cdot \vec{\sigma}_{\alpha, \beta} c_{\vec{k}, \beta} - \mu \, c\dg_{\vec{k}, \alpha} c_{\vec{k}, \alpha} \cr
H^{imp} & = & \sum_{\vec{k}, \vec{k'}} c\dg_{\vec{k}, \alpha} V_{\vec{k}, \vec{k}'} c_{\vec{k}', \alpha} 
\ea
Here, $V_{\vec{k}, \vec{k}'} = \sum_n V_n \me^{i n (\theta_{k} - \theta_{k'})}$, and $V_n \approx \tfrac{V a^2}{2} \frac{(k_F a)^n}{2^n \Gamma(\tfrac{n+1}{2})}$, while $v_F$ and $\sigma^i \in [\mI, \vec{\sigma}]$ are the Fermi velocity and spin Pauli matrices, and $k_F a$ determines $V_n$, which will be shown to control the skew scattering strength. Since the impurity is non-magnetic, the system is invariant under time-reversal symmetry, $\mathcal{T} = \mathcal{K} i \sigma_2$, $H= \mathcal{T} H \mathcal{T}^{-1}$. All the scattering events from an impurity are summed up in the $\Tbar$-matrix, and the spin-dependent skew scattering is captured by the $\sigma^{\pm}$ terms, illustrated in Fig.~\ref{fig: Fermi surface}. The following Dyson equations, in operator formalism, give the effective Green's function, $\hat{\Gbar}_{eff} = \hat{\Gbar}_0 + \hat{\Gbar}_0 \hat{\Tbar} \, \hat{\Gbar}_0$, and $\Tbar$-matrix, $\hat{\Tbar} = \hat{\Vbar} + \hat{\Vbar} \, \hat{\Gbar}_0 \hat{\Tbar}$, with $\hat{\Gbar}_0$ being the bare Green's function, and Fig.~\ref{fig: Geff Feyn diag} shows the Feynman diagram for the effective Green's function.
\begin{subequations}
\ba
\label{eqn: G0 & T-matrix}
\overline{\overline{G}}_0(\vec{k}, i \omega_n) & = & \frac{1}{i \omega_n + \mu - v_F \vec{k} \cdot \vec{\sigma}} \\
& = & g^0_0(k, i \omega_n) \mI + g^a_0(k, i \omega_n)(\cos \theta \, \sigma^x + \sin \theta \, \sigma^y) \cr
\Tbar(\vec{k}, \vec{k}', i \omega_n) & = & \sum_{nm} T^{i}_{nm}(|\vec{k}|, |\vec{k}'|, i \omega_n) \me^{i n \theta_{k}} \me^{-i m \theta_{k'}} \sigma^i 
\ea
\end{subequations}

Rotational symmetry of the Hamiltonian allows us to carry out a multipole expansion of $\overline{\overline{G}}_0(\vec{k}, i \omega_n)$ and the $\Tbar$-matrix, where $g^0_0(k, i \omega_n) = \tfrac{i \omega_n + \mu}{(i \omega_n + \mu)^2 - v_F^2 k^2}$, and $g^a_0(k, i \omega_n) = \tfrac{v_F k}{(i \omega_n + \mu)^2 - v_F^2 k^2}$. We assume the $\Tbar$-matrix varies slowly near $E_F$, i.e. absence of resonances, thereby simplifying the radial integral and reducing the Dyson equation to a set of coupled algebraic recurrence equations for the retarded $\Tbar$-matrix coefficients, $T^{i}_{nm}(|\vec{k}| = |\vec{k}'| = k_F, \omega = E_F)$.
\begin{subequations}
\ba
\label{eqn: T-reccurrence eqns}
T^{z \pm}_{nm} & = &  \delta_{n, m} \Big[V_n \bl 1 - V_{n \pm 1} \langle g_0^0(E_F) \rangle \br \Big] \Big[ \bl 1 - V_n \langle g_0^0(E_F) \rangle \br \cr
& & \times \bl 1 - V_{n \pm 1} \langle g_0^0(E_F) \rangle \br - V_n V_{n \pm 1} \langle g_0^1(E_F) \rangle^2 \Big]^{-1} \\
T^{\pm}_{nm} & = & \frac{\delta_{n \mp 1, m}}{2} \Big[ V_n V_{n \mp 1} \langle g_0^1(E_F) \rangle \Big] \Big[ \bl 1 - V_{n \mp 1} \langle g_0^0(E_F) \rangle \br \cr
 & & \times \bl 1 - V_{n} \langle g_0^0(E_F) \rangle \br - V_n V_{n \mp 1} \langle g_0^1(E_F) \rangle^2 \Big]^{-1}
\ea
\end{subequations}
The $\Tbar$-coefficients reduce to two set of coupled equations for $T^{z\pm} = T^0_{nm} \pm T^3_{nm}$ and $T^{\pm} = T^1_{nm} \pm i T^2_{nm}$, given in terms of $V_n$ and the momentum-averaged retarded Green's functions, $\langle g_0^{i, (R)}(\omega) \rangle = \int \tfrac{d k}{2 \pi} k g^{i, (R)}_0(k, \omega)$ (refer to SOM for calculation details). The arguments of the $\Tbar$-matrix coefficients are dropped, understanding that they are evaluated at $k_F$ and $E_F$. Defining the symmetric and asymmetric parts of the spin-flip scattering as $T^{S/A} = T^{+}_{10} \pm T^{-}_{-10}$, $T^3_0 \equiv T^3_{00}$, and $T^3_1 \equiv T^3_{11}$, we can now write down the $s$ and $p$-wave channels of the $\Tbar$-matrix.  
\ba
\label{eqn: T-matrix}
\Tbar(\theta_k, \theta_{k'}) & = & T^0 \mI + T^3_0 \sigma^z+ T^3_1 \big( \me^{i (\theta_k - \theta_{k'})} - \me^{-i (\theta_k - \theta_{k'})}\big) \sigma^z \cr
 & + & \frac{T^S + T^A}{2} \me^{i \theta_{k} } \sigma^{-} + \frac{T^S - T^A}{2} \me^{-i \theta_{k}} \sigma^{+}  \cr
 & + & \frac{T^S + T^A}{2} \me^{-i \theta_{k'} } \sigma^{+} + \frac{T^S - T^A}{2} \me^{i \theta_{k'}} \sigma^{-}
\ea
with detailed expressions for the $\Tbar$-matrix coefficients shown in the SOM. Charge-transport is dominated by the largest term, $|T^0| \propto V_0$, while spin-flip scatterings are captured by the $T^{S/A} \sigma^{\pm}$ terms. Upon projection into the upper helical band, we obtain a chiral spin-flip scattering term, $T^S \sin(\theta_{k} - \theta_{k'})$, which comes from $3^{rd}$ and higher orders in perturbation; $T^S \propto V_0 V_1^2 N_0(E_F)^2$, in agreement with previous work \cite{NagaosaSugimotoPRB2006}. Hence, the skew scattering strength can be tuned by varying $k_F a$, i.e. either the Fermi level or the impurity size $a$.

It is now straightforward to calculate the effective Green's function in the dilute impurity limit ($n_i \ll 1$)\cite{Rammer2004}, $\Gbar^{(R)}(\vec{k}, \omega) = \Big[ \omega - v_F \vec{k} \cdot \vec{\sigma} - \overline{\overline{\Sigma}}^{(R)}(\vec{k}, \omega)  \Big]^{-1}$, where the retarded self-energy is $\overline{\overline{\Sigma}}^{(R)}(\vec{k}, \omega) = n_i \sum_{\vec{k_1}} \Vbar(\vec{k}, \vec{k}_1) \Gbar^{(R)}_{eff}(\vec{k}_1, \omega) \Tbar^{(R)}(\vec{k}_1, \vec{k}, \omega)$. The appearance of $\Gbar^{(R)}_{eff}(\vec{k}, \omega)$ instead of $\Gbar^{(R)}_{0}(\vec{k}, \omega)$ reflects the presence of multiple impurities. We assume an average quasi-particle scattering rate near the Fermi surface, i.e. $\gbar \equiv Im[\overline{\overline{\Sigma}}^{(R)}(k_F, E_F)]$, and take $v_F$ and $E_F$ to be experimentally determined parameters, thereby dropping the real part of the self-energy.
\begin{subequations}
\ba
\label{eqn: qp scattering rate}
\gbar & = & \gamma_0 \mI - \gamma_a \bl \cos \theta \, \sigma^x + \sin \theta \, \sigma^y \br \cr
& - & \, \gamma_b (\sin \theta \, \sigma^x - \cos \theta \, \sigma^y) + i \gamma_3 \, \sigma^z \\
 \gamma_0 & = & n_i N^{(0)}_{eff}(E_F) \Big[ |T^0|^2 + |T^3_0|^2 \cr
 & & - 2\left( |T^3_1|^2 + |T^A|^2 - |T^S|^2 \right) \Big]  \\
 \gamma_a & = & 4 n_i N^{(1)}_{eff}(E_F) \left[ |T^S|^2 - |T^A|^2 \right] 
\ea
\end{subequations}

We have carried out a multipole expansion of $\gbar$, and the main quasi-particle scattering channels relevant to transport are the $s$ and $p$-wave $\gamma_0$ and $\gamma_a$ terms (refer to SOM for complete expressions of all $\gamma$). As we shall show later, the transport scattering rate, $\gamma_t$, will be given in terms of $\gamma_0$ and $\gamma_a$. The angular momentum resolved density of states (DOS) is defined as $N^{(i)}_{eff}(\omega) = \int \frac{k dk}{2 \pi} Im \left[ g^{i}_{eff}(k, \omega) \right]$, and $N^{(0)}_{eff}(E_F)$ and $N^{(1)}_{eff}(E_F)$ correspond to the $s$ and $p$-wave components respectively. Since scattering events that result in a change of angular momentum, i.e involving the $l = 1$ component $N^{(1)}_{eff}(E_F)$, will also cause a spin-flip due to spin-orbit coupling, we see that $\gamma_{0}$ and $\gamma_a$ are due to spin-independent and dependent scattering respectively.

The effective Green's function is therefore given by,
\ba
\label{eqn: eff Greens func}
\Gbar^{(R)}_{eff}(\vec{k}, \omega) & = & \left[ \omega + \mu - v_F \vec{k} \cdot \vec{\sigma} - i \gbar(\vec{k}, \omega) \right]^{-1} \\
& = & g^0_{eff}(k ,\omega) \mI + g^a_{eff}(k, \omega) \bl \cos \theta \, \sigma^x + \sin \theta \, \sigma^y \br \cr
 & + & g^b_{eff}(k, \omega) \bl \sin \theta \, \sigma^x - \cos \theta \, \sigma^y \br + g^3_{eff}(k, \omega) \sigma^z \nonumber
\ea
where,

\begin{subequations}
\ba
\label{eqn: g0 expr}
 g^0_{eff}(k, \omega) & = & \frac{(\Omega(k) + i \kappa(k)) (\omega + \mu - i \gamma_0 )}{\Omega^2(k) + \kappa^2(k)} \\
 \label{eqn: ga expr}
 g^a_{eff}(k, \omega) & = & \frac{(\Omega(k) + i \kappa(k)) (v_F |\vec{k}| + i \gamma_a )}{\Omega^2(k) + \kappa^2(k)}
\ea
\end{subequations}
with $\Omega(k) = (\omega + \mu)^2 - v_F^2 |\vec{k}|^2 - \gamma_0^2 + \gamma_a^2 + \gamma_b^2 - \gamma_3^2$, and $\kappa(k) = 2 \bl (\omega + \mu) \gamma_0 + v_F |\vec{k}| \gamma_a \br$.

A similar multipole expansion of $\Gbar^{(R)}_{eff}(\vec{k}, \omega)$ has been done, and we show here only the main $s$ and $p$-wave terms, $g^0_{eff}(k, \omega)$ and $ g^a_{eff}(k, \omega)$, with complete expressions for the scattering-induced $g^b_{eff}(k, \omega)$ and $g^3_{eff}(k, \omega)$ terms relegated to the SOM for brevity. From Eqs.~(\ref{eqn: g0 expr}) \&~(\ref{eqn: ga expr}), it is clear that Weyl fermions in the $s$ and $p$-wave channels pick up a $\gamma_0$ and $\gamma_a$ scattering rate respectively, and we shall show later that it is chiral scattering between the $s$ and $p$-wave electrons that drive the SHE.

$\Gbar^{(R)}_{eff}(\vec{k}, \omega)$ and $\overline{\overline{\Sigma}}^{(R)}(\vec{k}, \omega)$ are determined self-consistently by solving Eqns.~\ref{eqn: qp scattering rate} \& \ref{eqn: eff Greens func}, i.e. $\overline{\overline{\Sigma}}^{(R)}(\vec{k}, \omega)$ is calculated using the disorder-averaged density of states, $N^{(i)}_{eff}(\omega) = \int \frac{k dk}{2 \pi} Im \left[ g^{i}_{eff}(k, \omega) \right]$. However, in the dilute impurity limit, $N^{(0)/(1)}_{eff}(E_F) = \tfrac{N_{0}(E_F)}{2} (1 + O(\gbar))$ \cite{Rammer2004}; allowing us to drop the $O(n_i)$ corrections.

As stated earlier, the DC longitudinal charge conductivity, spin-Hall conductivity and spin accumulation are given by analytic continuation of the corresponding Matsubara correlation functions, 
\begin{subequations}
\ba
\label{eqn: pi_yy}
\pi_{yy}(\vec{k}, i \omega_n) & = & - \int^{\beta}_{0} d \tau \me^{-i \omega_n \tau} \langle T_{\tau } \, j_{y}(\vec{k}, \tau) j_{y}(\vec{k}, 0)  \rangle \\
\label{eqn: pi_iy}
\pi^{i}_{y}(\vec{k}, i \omega_n) & = & - \int^{\beta}_{0} d \tau \me^{-i \omega_n \tau} \langle T_{\tau} \, \sigma^i(\vec{k}, \tau) j_{y}(\vec{k}, 0)  \rangle \\
\label{eqn: pi_zxy}
\pi^{z}_{xy}(\vec{k}, i \omega_n) & = & -\int^{\beta}_{0} d \tau \me^{-i \omega_n \tau} \langle T_{\tau} \, P^{z}_{x}(\vec{k}, \tau) j_{y}(\vec{k}, 0)  \rangle
\ea
\end{subequations}
Note that $\pi_{yy}$ and $\pi^{i}_{y}$ are equal up to a factor of $\tfrac{e v_F}{\hbar}$ for Weyl fermions due to spin-momentum locking, i.e. $\hat{j_y} = e v_F \hat{\sigma}^y$. The spin torque current, $P^{z}_x$, arises from the intrinsic quantum-mechanical evolution of the electron spin, and the $z$-component of the spin-torque current along $\hat{x}$ is,
\ba
\label{eqn: spin torque}
P^{z}_x(\vec{k}) & = & \tfrac{i}{k_x} \tfrac{d \hat{S}^z(\vec{k})}{d t} \\
 & = &\frac{2 v_F}{i k_x} \sum_{\vec{p}} c\dg_{\vec{p}, \sigma} \left[  \left( \vec{p} + \frac{\vec{k}}{2} \right)_{y}\sigma^x - \left( \vec{p} + \frac{\vec{k}}{2} \right)_{x} \sigma^y  \right] c_{\vec{p} + \vec{k}, \sigma'} \nonumber
\ea

\begin{figure}[bht!]
\begin{center}
\includegraphics[trim=0mm 0mm 0mm 0mm, clip, width=\columnwidth]{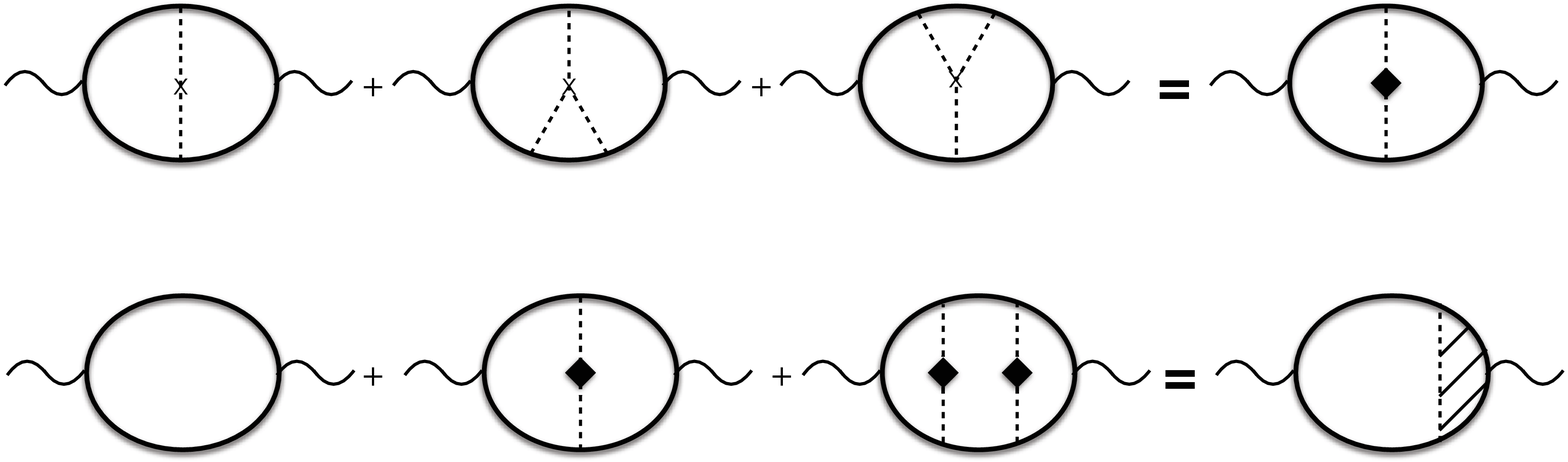}
\end{center}
\captionsetup{justification=RaggedRight, singlelinecheck=false}
\caption{\label{fig: Vertex Feyn diag} 
Feynman diagram for the effective scattering vertex, $\Gammabar^{y}(\vec{p}, \omega)$, is shown in the second line. This includes an infinite subset of scattering events from the dilute concentration of  impurities. The first line shows all the scattering events from a single impurity, and the second and third diagrams in the first line are the leading-order contributions to skew scattering.}
\end{figure}

The Feynman diagrams for these correlation functions are shown in Fig.~\ref{fig: Vertex Feyn diag}, with chiral spin-flip scattering starting to contribute at third-order in perturbation theory. Fig.~\ref{fig: Vertex Feyn diag} shows the infinite subset of Feynman ladder diagrams summed up in the Bethe Salpeter equation for the scattering vertex,
\begin{widetext}
\ba
\label{eqn: vertex Dyson eqn}
\Gammabar^{y}(\vec{k} + \vec{p}, \vec{p}, i \Omega_m + i \omega_n, i \omega_n) = \sigma^y + \sum_{\vec{q}} \Tbar(\vec{k} + \vec{p}, \vec{k} + \vec{q}, i \Omega_m + i \omega_n) \Gbar_{eff}(\vec{k}+\vec{q}, i \Omega_m + i \omega_n) \cr
 \times \Gammabar^{y}(\vec{k} + \vec{q}, \vec{q}, i \Omega_m + i \omega_n, i \omega_n) \Gbar_{eff}(\vec{q}, i \omega_n) \Tbar(\vec{q}, \vec{p}, i \omega_n)
\ea
\end{widetext}

Here, $\vec{k}$ and $i \Omega_m$ are the  external momentum and frequency, and the uniform DC limit of the conductivities is obtained by analytic continuation of $i \Omega_m \rightarrow \omega + i \eta$, and taking the limit $\vec{k} \rightarrow 0$ followed by $\omega \rightarrow 0$. Hence, we only need to calculate the on-shell component of the scattering vertex $\Gammabar^{y}(\vec{p}, \omega) = \Gammabar^{y}(\vec{p}, \omega - i \eta, \omega + i \eta)$. The Bethe-Salpeter equation for $\Gammabar^{y}(\vec{p}, \omega)$ is solved self-consistently by expanding $\Gammabar^{y}(\vec{p}, \omega) = \sum_{n} \Gamma^{i}_{n} \me^{i n \theta} \sigma^i $ in multipole terms, assuming that the $\Tbar$-matrix and $\Gammabar^{y}(\vec{p}, \omega)$ vary slowly near $E_F$ (see SOM for details). Keeping only the $s$- and $p$-wave channels, and evaluating $\Gammabar^{y}(|\vec{p}| = k_F, \omega = E_F)$ at the Fermi surface, we obtain,
\ba
\label{eqn: vertex func}
\Gammabar^{y}(k_F, E_F) & = & (\Gamma^0_{p_x} \cos \theta + i \Gamma^0_{p_y} \sin \theta) \, \mI \\
 & & + \Gamma^1_0(E_F) \sigma^x + \Gamma^2_0(E_F) \, \sigma^y  \cr
 & & + (\Gamma^3_{p_x}(E_F) \cos \theta + i \Gamma^3_{p_y}(E_F) \sin \theta) \, \sigma^z \nonumber
\ea 
where,
\be
\Gamma^2_0 = \frac{\gamma_0}{\gamma_{t}}  \text{,}    \hspace{2cm}   \Gamma^3_{p_x} = - \tfrac{\gamma_{s}}{\gamma_t}.
\ee 

After analytic continuation of the current-current correlation functions in Eq.~(\ref{eqn: pi_yy})  - (\ref{eqn: pi_zxy}), we find that the main contributions come from the $\Gamma^2_0$ charge-transport and $\Gamma^3_{p_x}$ spin-transport scattering vertices (refer to SOM for all the $\Gammabar$-coefficients). We can therefore define a transport and chiral spin-flip scattering rate respectively as, 
\be
\label{eqn: transport coeffs}
\gamma_t = ( \tfrac{1}{2} \gamma_0 + \gamma_a) \text{,}    \hspace{1cm}  \gamma_{s} = \frac{n_i \pi N_0(E_F)}{2} |T^0| |T^S|.
\ee

The main results of this paper are the charge and spin conductivities, and the Rashba-Edelstein coefficient,
\begin{subequations}
\ba
\label{eqn: sigmayy}
 \sigma_{yy} & = & \bl e v_F \br^2 \frac{N_0(E_F)}{2} \frac{1}{\gamma_{t}}  \\
\label{eqn: sigmazxy}
 \sigma^{z}_{xy} & = & - \hbar e v_F^2 \frac{N_0(E_F)}{2} \frac{1}{\gamma_{t}}  \frac{\gamma_{s}}{\gamma_0 + \gamma_a} \\
 \label{eqn: sigmaiy}
\sigma^{y}_{y} & = & \hbar e v_F \frac{N_0(E_F)}{2} \frac{1}{\gamma_{t}} 
\ea
\end{subequations}

Our key finding is Eq.~(\ref{eqn: sigmazxy}), which shows an $O(\tfrac{1}{n_i})$ skew scattering contribution to the SHE. Explicitly writing out the spin and angular-momentum scattering channels for $\sigma^{z}_{xy} = \hbar e v_F^2 Re[\Gamma^3_{p_x} (\eta^{a0}(E_F) - \eta^{0a}(E_F))]$, where $\eta^{ij}(\omega) = \int \tfrac{d p}{2 \pi} p^2 \tfrac{\partial g^{i(R)}_{eff}(p, \omega)}{\partial p} g^{j(A)}_{eff}(p, \omega)$, we see that chiral spin-flip scattering between the $s$ and $p$-wave electrons is the cause of the skew-scattering mechanism, and the strength of which is measured via the spin-Hall angle, 
\ba
 \theta^{SH}  & = & -\frac{\hbar}{e} \frac{\gamma_{s}}{\gamma_0 + \gamma_a} 
\ea
Here, $e < 0$ is the electron charge, and power counting of  $\gamma_t \sim \gamma_0 \sim n_i V_0^2 N_0(E_F)$ and $\gamma_{s} \sim n_i V_0^2 V_1^2 N_0(E_F)^3$, gives $\theta^{SH} \sim O\left( \tfrac{V^2}{v_F^2/a^2} (k_F a)^4 \right)$. This is our key result: $\theta^{SH}$ has a fixed positive sign, and is a strongly-dependent function of $k_F a$; hence, the SHE can be tuned by $E_F$.

Finally, we briefly discuss the effects of band bending in Weyl systems. The leading $O(\tfrac{1}{m})$ correction comes from including a conventional spin current, $\vec{j}^z_{s} = \psi\dg \tfrac{1}{2} \{\vec{v}, S^z \} \psi$, with $\vec{v} = \tfrac{\hbar \vec{k}}{m}$. However, it has been pointed out\cite{MolenkampPRB2003, HalperinMischenkoPRL2004, NagaosaSugimotoPRB2006} that $\vec{j}^z_{s} \propto \dot{\sigma_y}$ for Rashba-type systems; hence, up to $O(\tfrac{1}{m})$, band bending does not give rise to a spin current for Weyl fermion systems.

In conclusion, we have analysed both the spin Hall and Rashba-Edelstein effects in a 2D Weyl electron system. Our results show that strong spin-orbit coupling in the band-structure is sufficient to cause chiral spin-flip scattering of the helical electrons off non-SOC scalar impurities, resulting in a skew-scattering contribution to the SHE. The strength of this mechanism is measured by the SHA, $\theta^{SH} = -\tfrac{\hbar}{e} \frac{\gamma_{s}}{\gamma_0 + \gamma_a} \sim -\tfrac{\hbar}{e} \; O\left( \tfrac{V^2}{v_F^2/a^2} (k_F a)^4 \right) $, and we highlight the fact that the skew scattering strength can be tuned by varying $k_F a$, thereby providing an experimentally-accessible parameter for controlling the SHE. In addition, we have also found an $O(\tfrac{1}{\gamma_t})$ Rashba-Edelstein effect due to spin-momentum locking of the Weyl fermions. We gratefully acknowledge I. Mertig, K. Kondou and Y. Tokura for helpful discussions, and this work was supported by CREST, Japan Science and Technology Agency (JST).

\bibliography{main.bib}

\end{document}


\centerline{\bf
Supplementary Online Material: Spin Hall Effect on Topological Insulator Surface}

\author{T.~Tzen Ong}
\affiliation{RIKEN Center for Emergent Matter Science (CEMS), Saitama 351-0198, Japan}
\affiliation{Department of Applied Physics, University of Tokyo, Tokyo 113-8656, Japan}
\date{\today}

\tableofcontents

\section{2D Weyl Fermion and Chiral Skew Scattering from Non-magnetic Impurity}

We consider elastic scattering near $E_F$ of 2D Weyl fermions (Dresselhaus-type $v_F \, \vec{k} \cdot \vec{\sigma}$ system) from a dilute ($n_i \ll 1$) random distribution of non-magnetic impurities, at positions $\vec{R}_i$, with impurity scattering $H^{imp} = \sum_{\vec{r}, \vec{R}_i} V \mathrm{e}^{-\frac{|\vec{r}-\vec{R}_i|^2}{a^2}} c\dg_{\sigma}(\vec{r})  \mI_{\sigma \sigma'} c_{\sigma'}(\vec{r})$, and the impurity size $a$ determines the strength of skew scattering. Note that the results can be easily translated into the Rashba-type $v_F \, \hat{z} \times \vec{k} \cdot \vec{\sigma}$ case by rotating the momentum by $90^{\degree}$. The chemical potential $\mu$ is chosen to lie in the upper helical band, with the upper/ lower helical Weyl fermions being $\psi_{\pm, \vec{k}} = \tfrac{1}{\sqrt{2}} (\pm \, c_{\vec{k}, \uparrow} + \me^{i \theta_k} c_{\vec{k}, \downarrow})$,  and the Hamiltonian is,
\ba
\label{eqn: Hamiltonian}
H & = & H^0 + H^{imp} \\
H^0 & = & \sum_{\vec{k}, \alpha, \beta} c\dg_{\vec{k}, \alpha}  v_F  \vec{k} \cdot \vec{\sigma}_{\alpha, \beta} c_{\vec{k}, \beta} - \mu \, c\dg_{\vec{k}, \alpha} c_{\vec{k}, \alpha} \cr
H^{imp} & = & \sum_{\vec{k}, \vec{k'}} c\dg_{\vec{k}, \alpha} V_{\vec{k}, \vec{k}', \alpha \beta} c_{\vec{k}', \beta} 
\ea

The non-magnetic impurity is modelled with a scattering potential $V$ and a Gaussian profile, $V \, \me^{-\frac{r^2}{a^2}}$. Hence the scattering matrix element of 2D Weyl fermions off this impurity is,
\ba
V_{\vec{k}, \vec{k}', \sigma \sigma'} & = & \bra{\vec{k}, \sigma} V \me^{-\frac{r^2}{a^2}} \ket{\vec{k}', \sigma'} \cr
 & = & \sum_n V_n \me^{i n (\theta_{k} - \theta_{k'})} \mI_{\sigma \sigma'}
 \ea 
where $V_n = \tfrac{V a^2}{8} \me^{-\frac{1}{8} k_F^2 a^2} k_F a \Big( I(\tfrac{n-1}{2}, \tfrac{k_F^2 a^2}{8}) - I(\tfrac{n+1}{2}, \tfrac{k_F^2 a^2}{8}) \Big) \approx \tfrac{V a^2}{2} \frac{(k_F a)^n}{2^n \Gamma(\tfrac{n+1}{2})}$. We have assumed that transport involves mainly the quasi-particles near $E_F$, i.e. $|\vec{k}| = |\vec{k}'| \approx k_F$, and have used the result $\int_{0}^{\infty} r dr J_n(k_F r) \me^{-\tfrac{r^2}{a^2}} =  \tfrac{a^2}{8}  k_F a \, \me^{-\frac{1}{8} k_F^2 a^2} \Big( I(\tfrac{n-1}{2}, \tfrac{k_F^2 a^2}{8}) - I(\tfrac{n+1}{2}, \tfrac{k_F^2 a^2}{8}) \Big)$, with $J(n,z)$ and $I(n,z)$ being the Bessel and modified Bessel functions of the first kind respectively, and $\Gamma(n)$ is the Gamma function

All the scattering events from a single impurity are captured in the $\Tbar$-matrix, given by the Dyson equation $\hat{\Tbar} = \hat{\Vbar} + \hat{\Vbar} \, \hat{\Gbar}_{0} \, \hat{\Tbar}$. Making use of the rotational symmetry of the system, we express the Greens function and $\Tbar$-matrix in a multipole-expansion,
\ba
\overline{\overline{G}}_0(\vec{k}, i \omega_n) & = & \frac{1}{i \omega_n + \mu - v_F \vec{k} \cdot \vec{\sigma}} \\
& = & g^0_0(k, i \omega_n) \mI + g^1_0(k, i \omega_n)(\cos \theta_k \, \sigma^x + \sin \theta_k \, \sigma^y) \hspace{1cm} \text{, where} \cr
g^{0}_{0}(k, i \omega_n) & = & \frac{i \omega_n + \mu}{(i \omega_n + \mu)^2 - v_F^2 k^2} \cr
g^{1}_{0}(k, i \omega_n) & = & \frac{v_F k}{(i \omega_n + \mu)^2 - v_F^2 k^2} \nonumber
\ea
\ba
\label{eqn: T matrix}
\Tbar(\vec{k}, \vec{k}') & \equiv & \sum_{nm} T^{i}_{nm} \me^{i n \theta_k} \me^{-i m \theta_{k'}} \sigma^i \cr 
 & = & \Vbar({\vec{k}, \vec{k}'})  + \sum_{n_1 n_2 n_3}  \int \frac{d \theta_{k_1}}{2 \pi} \int \frac{k_1 d k_1}{2 \pi} V_{n_1} \me^{i n_1 (\theta_{k} - \theta_{k_1})} \cr
 & & \times \left[ g^0_0(k_1, i \omega_n) \mI + g^1_0(k_1, i \omega_n)(\cos \theta_{k_1} \, \sigma^x + \sin \theta_{k_1} \, \sigma^y) \right] \cr
 & & \times T^{j}_{n_2 n_3}(k_1, k') \me^{i n_2 \theta_{k_1}} \me^{-i n_3 \theta_{k'}} \sigma^j
 \ea 
The Pauli matrices are defined as $\sigma^i \in [\mI, \vec{\sigma}]$. As discussed in the main paper, we shall assume that there are no resonances, so the $\Tbar$-matrix varies slowly as a function of $\vec{k}$ near $E_F$. Approximating the $\Tbar$-matrix as a constant near $k_F$, the $\int dk_1$-integral is carried out only over the Green's function. This is the momentum-averaged retarded Green's function, $\langle g^{i,(R, A)}_0(i \omega_n) \rangle \equiv \int \frac{k dk}{2 \pi} g^{i,(R, A)}(k, i \omega_n) $, and the results are,
\begin{subequations}
\ba
\label{eqn: momentum avg Greens func}
\langle g^{0,(R, A)}_0(E_F) \rangle & = &  \mp \frac{i \pi}{2} N_0(E_F) sgn(E_F)  \\
\langle g^{1,(R, A)}_0(E_F) \rangle & = & \pm \frac{i \pi}{2} N_0(E_F) sgn(E_F)   
\ea
\end{subequations}

Here, $N_0(E_F) = \tfrac{E_F}{2 \pi v_F^2}$ is the bare density of states, and in terms of the momentum-averaged retarded Greens functions, the retarded $\Tbar$-matrix is now given by,

\ba
\label{eqn: T matrix Dyson eqn}
\Tbar(\vec{k}, \vec{k}')  & = & \sum_{n m} V_n \me^{i n (\theta_{k} - \theta_{k'})} \mI \delta_{nm} + V_n \me^{i n \theta_k} \me^{- i m \theta_{k'}} \Big[ \langle g^0_0(E_F) \rangle \big( T^{0}_{n m} \mI + T^{1}_{n m} \sigma^x + T^{2}_{n m} \sigma^y + T^{3}_{n m} \sigma^z \big) \\
 & + & \langle g^1_0(E_F) \rangle \big( T^{-}_{n-1 m} \mI + T^{-}_{n-1 m} \sigma^z + T^{z+}_{n-1 m} \sigma^- \big) + \langle g^1_0(E_F) \rangle \big( T^{+}_{n+1 m} \mI + T^{+}_{n+1 m} \sigma^z + T^{z-}_{n+1 m} \sigma^+ \big) \Big] \nonumber
\ea

The coefficients of the $\Tbar$-matrix are $T^{z\pm}_{nm} \equiv T^{0}_{nm} \pm T^{3}_{nm}$, $T^{\pm}_{nm} \equiv \tfrac{1}{2} \bl T^{1}_{nm} \pm i T^{2}_{nm} \br$, and are now defined by the following set of coupled recurrence equations,
\ba
\label{eqn: T-reccurrence eqns}
T^{z+}_{nm} & = & V_n \delta_{nm}  + V_n \langle g^0(E_F) \rangle T^{z+}_{nm} + 2 V_{n} \langle g^1(E_F) \rangle T^{+}_{n+1 m} \cr
T^{+}_{nm} & = & V_n \langle g^0(E_F) \rangle T^{+}_{nm} + \frac{1}{2} V_{n} \langle g^1(E_F) \rangle T^{z+}_{n-1 m} \cr
T^{z-}_{nm} & = & V_n \delta_{nm}  + V_n \langle g^0(E_F) \rangle T^{z-}_{nm} + 2 V_{n} \langle g^1(E_F) \rangle T^{-}_{n-1 m} \cr
T^{-}_{nm} & = & V_n \langle g^0(E_F) \rangle T^{-}_{nm} + \frac{1}{2} V_{n} \langle g^1(E_F) \rangle T^{z-}_{n+1 m}
\ea  

The $\Tbar$-coefficients reduce to two set of coupled equations for $T^{z\pm} = T^0_{nm} \pm T^3_{nm}$ and $T^{\pm} = T^1_{nm} \pm i T^2_{nm}$, given in terms of $V_n$ and the momentum-averaged retarded Green's functions, $\langle g^{i, (R)}(E_F) \rangle$. The arguments of the $\Tbar$-matrix coefficients are dropped, understanding that they are evaluated at $k_F$ and $E_F$. Some straightforward, albeit tedious, algebra allows us to solve Eq.~\ref{eqn: T-reccurrence eqns}.
\ba
T^{z+}_{nm} & = & \frac{V_n \bl 1 - V_{n+1} \langle g^0(E_F \rangle) \br \delta_{nm}}{\bl 1 - V_n \langle g^0(E_F) \rangle \br \bl 1 - V_{n+1} \langle g^0(E_F) \rangle \br - V_n V_{n+1} \langle g^1(E_F) \rangle^2} \cr
T^{+}_{nm} & = & \frac{1}{2} \frac{V_n V_{n-1} \langle g^1(E_F \rangle) \delta_{n-1 m} }{\bl 1 - V_{n-1} \langle g^0(E_F) \rangle \br \bl 1 - V_{n} \langle g^0(E_F) \rangle \br - V_n V_{n-1} \langle g^1(E_F) \rangle^2} \cr
T^{z-}_{nm} & = & \frac{V_n \bl 1 - V_{n-1} \langle g^0(E_F \rangle) \br \delta_{nm} }{\bl 1 - V_n \langle g^0(E_F) \rangle \br \bl 1 - V_{n-1} \langle g^0(E_F) \rangle \br - V_n V_{n-1} \langle g^1(E_F) \rangle^2} \cr
T^{-}_{nm} & = & \frac{1}{2} \frac{V_n V_{n+1} \langle g^1(E_F \rangle) \delta_{n+1 m} }{\bl 1 - V_{n+1} \langle g^0(E_F) \rangle \br \bl 1 - V_{n} \langle g^0(E_F) \rangle \br - V_n V_{n+1} \langle g^1(E_F) \rangle^2}
\ea
Therefore, the $\Tbar$-matrix coefficients are,
\ba
T^{0}_{nm} & = & \frac{1}{2} \frac{V_n \bl 1 - V_{n+1} \langle g^0(E_F \rangle) \br \delta_{nm}}{\bl 1 - V_n \langle g^0(E_F) \rangle \br \bl 1 - V_{n+1} \langle g^0(E_F) \rangle \br - V_n V_{n+1} \langle g^1(E_F) \rangle^2} \cr
 & + & \frac{1}{2} \frac{V_n \bl 1 - V_{n-1} \langle g^0(E_F \rangle) \br \delta_{nm} }{\bl 1 - V_n \langle g^0(E_F) \rangle \br \bl 1 - V_{n-1} \langle g^0(E_F) \rangle \br - V_n V_{n-1} \langle g^1(E_F) \rangle^2}\cr
T^{3}_{nm} & = & \frac{1}{2} \frac{V_n \bl 1 - V_{n+1} \langle g^0(E_F \rangle) \br \delta_{nm}}{\bl 1 - V_n \langle g^0(E_F) \rangle \br \bl 1 - V_{n+1} \langle g^0(E_F) \rangle \br - V_n V_{n+1} \langle g^1(E_F) \rangle^2} \cr
& - &  \frac{1}{2} \frac{V_n \bl 1 - V_{n-1} \langle g^0(E_F \rangle) \br \delta_{nm} }{\bl 1 - V_n \langle g^0(E_F) \rangle \br \bl 1 - V_{n-1} \langle g^0(E_F) \rangle \br - V_n V_{n-1} \langle g^1(E_F) \rangle^2}\cr
T^{1}_{nm} & = & \frac{1}{2} \frac{V_n V_{n-1} \langle g^1(E_F \rangle) \delta_{n-1 m} }{\bl 1 - V_{n-1} \langle g^0(E_F) \rangle \br \bl 1 - V_{n} \langle g^0(E_F) \rangle \br - V_n V_{n-1} \langle g^1(E_F) \rangle^2} \cr
& + & \frac{1}{2} \frac{V_n V_{n+1} \langle g^1(E_F \rangle) \delta_{n+1 m} }{\bl 1 - V_{n+1} \langle g^0(E_F) \rangle \br \bl 1 - V_{n} \langle g^0(E_F) \rangle \br - V_n V_{n+1} \langle g^1(E_F) \rangle^2} \cr
T^{2}_{nm} & = & -\frac{i}{2} \frac{V_n V_{n-1} \langle g^1(E_F \rangle) \delta_{n-1 m} }{\bl 1 - V_{n-1} \langle g^0(E_F) \rangle \br \bl 1 - V_{n} \langle g^0(E_F) \rangle \br - V_n V_{n-1} \langle g^1(E_F) \rangle^2} \cr
& + & \frac{i}{2} \frac{V_n V_{n+1} \langle g^1(E_F \rangle) \delta_{n+1 m} }{\bl 1 - V_{n+1} \langle g^0(E_F) \rangle \br \bl 1 - V_{n} \langle g^0(E_F) \rangle \br - V_n V_{n+1} \langle g^1(E_F) \rangle^2}
\ea
We calculate the $\Tbar$-matrix up to order $O(V_0 V_1^2)$, at which skew scattering appears, and keep only the $l = 0$ and $l = 1$ channels. Defining the symmetric and asymmetric parts of the spin-flip scattering as $T^{S/A} = T^{+}_{10} \pm T^{-}_{-10}$, we can now write down the $s$ and $p$-wave channels of the $\Tbar$-matrix.
\ba
\label{eqn: T-matrix}
\Tbar(\theta_k, \theta_{k'}) & = & T^0 \mI + T^3_0 \sigma^z+ T^3_1 \big( \me^{i (\theta_k - \theta_{k'})} - \me^{-i (\theta_k - \theta_{k'})}\big) \sigma^z \cr
 & + & \frac{T^S + T^A}{2} \me^{i \theta_{k} } \sigma^{-} + \frac{T^S - T^A}{2} \me^{-i \theta_{k}} \sigma^{+}  \cr
 & + & \frac{T^S + T^A}{2} \me^{-i \theta_{k'} } \sigma^{+} + \frac{T^S - T^A}{2} \me^{i \theta_{k'}} \sigma^{-}
\ea
and the coefficients are defined as,
\begin{subequations}
\ba
\label{eqn: T-coefficients}
T^0 & = & \frac{1}{2} \frac{V_0 \big( 1- V_1 \langle g^0(E_F) \rangle \big) }{\Big[ \bl 1 - V_0 \langle g^0(E_F) \rangle \br \bl 1 - V_1 \langle g^0(E_F) \rangle \br - V_0 V_{1} \langle g^1(E_F) \rangle^2 \Big]} \cr
 & & + \frac{1}{2} \frac{V_0 \big( 1- V_{-1} \langle g^0(E_F) \rangle \big) }{\Big[ \bl 1 - V_0 \langle g^0(E_F) \rangle \br \bl 1 - V_{-1} \langle g^0(E_F) \rangle \br - V_0 V_{-1} \langle g^1(E_F) \rangle^2 \Big]} \cr
 & = & \frac{V_0}{\Big[ 1 - V_0 \langle g^0(E_F)\rangle \Big]^2} \\
T^3_0 & = & \frac{1}{2} \frac{V_0 \big( 1- V_1 \langle g^0(E_F) \rangle \big) }{\Big[ \bl 1 - V_0 \langle g^0(E_F) \rangle \br \bl 1 - V_1 \langle g^0(E_F) \rangle \br - V_0 V_{1} \langle g^1(E_F) \rangle^2 \Big]} \cr
 & & - \frac{1}{2} \frac{V_0 \big( 1- V_{-1} \langle g^0(E_F) \rangle \big) }{\Big[ \bl 1 - V_0 \langle g^0(E_F) \rangle \br \bl 1 - V_{1} \langle g^0(E_F) \rangle \br - V_0 V_{-1} \langle g^1(E_F) \rangle^2 \Big]} \cr
 & = & \frac{V_0^2 V_1 \langle g^{1}(E_F) \rangle^2}{\Big[ 1 - V_0 \langle g^0(E_F)\rangle  \Big]^2} \\
 T^3_1 & = & \frac{1}{2} \frac{V_1 \big( 1- V_2 \langle g^0(E_F) \rangle \big) }{\Big[ \bl 1 - V_1 \langle g^0(E_F) \rangle \br \bl 1 - V_2 \langle g^0(E_F) \rangle \br - V_1 V_{2} \langle g^1(E_F) \rangle^2 \Big]} \cr
 & & - \frac{1}{2} \frac{V_1 \big( 1- V_{0} \langle g^0(E_F) \rangle \big) }{\Big[ \bl 1 - V_1 \langle g^0(E_F) \rangle \br \bl 1 - V_{0} \langle g^0(E_F) \rangle \br - V_1 V_{0} \langle g^1(E_F) \rangle^2 \Big]} \cr
 & = & -\frac{1}{2} \frac{V_0 V_1^2 \langle g^{1}(E_F) \rangle^2}{\Big[ 1 - V_1 \langle g^0(E_F)\rangle  \Big]^2} \\
T^S & = & \frac{1}{2} \frac{V_0 V_1 \langle g^1(E_F) \rangle}{(1 - V_0 \langle g^0(E_F) \rangle )(1 - V_1 \langle g^0(E_F) \rangle ) - V_0 V_1 \langle g^1(E_F) \rangle^2} \cr
& & + \frac{1}{2} \frac{V_0 V_{-1} \langle g^1(E_F) \rangle}{(1 - V_0 \langle g^0(E_F) \rangle )(1 - V_{-1} \langle g^0(E_F) \rangle ) - V_0 V_{-1} \langle g^1(E_F) \rangle^2} \cr
& = & \frac{V_0 V_1^2 \langle g^0(E_F) \rangle \langle g^1(E_F) \rangle }{\Big[ 1 - V_0 \langle g^0(E_F) \rangle \Big]^2} \\
T^A & = & \frac{1}{2} \frac{V_0 V_1 \langle g^1(E_F) \rangle}{(1 - V_0 \langle g^0(E_F) \rangle )(1 - V_1 \langle g^0(E_F) \rangle ) - V_0 V_1 \langle g^1(E_F) \rangle^2} \cr
& & - \frac{1}{2} \frac{V_0 V_{-1} \langle g^1(E_F) \rangle}{(1 - V_0 \langle g^0(E_F) \rangle )(1 - V_{-1} \langle g^0(E_F) \rangle ) - V_0 V_{-1} \langle g^1(E_F) \rangle^2} \cr
& = & \frac{V_0 V_1 \langle g^1(E_F) \rangle }{\Big[ 1 - V_0 \langle g^0(E_F) \rangle \Big]^2}
\ea
\end{subequations}

We point out that upon projecting into the upper helical band, i.e. calculating the matrix elements $\bra{\vec{k}, +} T^S (\me^{i \theta_{k}} \sigma^- + \me^{-i \theta_{k}'} \sigma^+) \ket{\vec{k}', +} = 2 \, T^S \left(  \cos(\theta_{k} - \theta_{k}') -i \sin(\theta_{k} - \theta_{k}') \right)$, we find that the spin-flip scattering gives rise to a skew-scattering term $2 i T^S \sin(\theta_{k} - \theta_{k}')$ in the chiral band basis, which will drive the SHE. 

\section{Effective Greens Function and Quasi-particle Scattering Rate}

The retarded $\Tbar$-matrix calculated in Eq.~\ref{eqn: T matrix} includes only scattering from a single impurity, and in the dilute impurity limit, the $\Tbar$-matrix for scattering from all impurities can be calculated in the non-crossing approximation NCA)~\cite{Rammer2004}) by including scattering events from other impurities in the bare Greens function leg, i.e. replacing $\Gbar_0$ by $\Gbar_{eff}$, in the calculation of the $\Tbar$-matrix. Hence, this forms an implicit self-consistent solution for the retarded and advanced $\Gbar_{eff}$ function and $\Tbar$-matrix.
\ba
\label{eqn: dilute limit T matrix}
\Tbar^{(R)}(\vec{k}, \vec{k}') & = & n_i \Vbar(\vec{k}, \vec{k}') + n_i \sum_{\vec{k}_1} \Vbar(\vec{k}, \vec{k}_1) \Gbar^{(R)}_{eff}(\vec{k}_1, \omega) \Tbar^{(R)}(\vec{k}_1, \vec{k}',\omega) \cr
\Tbar^{(A)}(\vec{k}, \vec{k}') & = & n_i \Vbar(\vec{k}, \vec{k}') + n_i \sum_{\vec{k}_1} \Vbar(\vec{k}, \vec{k}_1) \Gbar^{(A)}_{eff}(\vec{k}_1, \omega) \Tbar^{(A)}(\vec{k}_1, \vec{k}',\omega)
\ea
In the non-crossing approximation, the retarded self-energy $\overline{\overline{\Sigma}}^{(R)}(\vec{k}, \omega)$ and quasi-particle scattering rate $\gbar(\vec{k}, \omega) = Im[\overline{\overline{\Sigma}}^{(R)}(\vec{k}, \omega)]$ are given by,
\ba
\label{eqn: Self energy}
\overline{\overline{\Sigma}}^{(R)}(\vec{k}, \omega) & = & n_i \sum_{\vec{k}_1} \Vbar(\vec{k}, \vec{k}_1) \Gbar^{(R)}_{eff}(\vec{k}_1, \omega) \Tbar^{(R)}(\vec{k}_1, \vec{k}',\omega) \cr
\gbar(\vec{k}, \omega) = Im[\overline{\overline{\Sigma}}^{(R)}(\vec{k}, \omega)] & = & \sum_{\vec{k}_1} \Tbar^{(A)}(\vec{k}, \vec{k}_1, \omega) \overline{\overline{A}}_{eff}(\vec{k}_1, \omega) \Tbar^{(R)}(\vec{k}_1, \vec{k}, \omega)
\ea
The spin-dependent spectral weight is given by $\overline{\overline{A}}_{eff}(\vec{k}, \omega) = 2 Im[\Gbar_{eff}^{(R)}(\vec{k}, \omega)]$. Similar to the calculation of the $\Tbar$-matrix, the $\int dk$-integral for the self-energy is done using the approximation that the $\Tbar$-matrix varies slowly near $k_F$, leaving only the $\int dk$-integral of the spin-dependent spectral weight, which is none other than the density of states,
\ba
\label{eqn: DOS}
N_{eff}^{(0)}(\omega) & = & \int \frac{k dk}{2 \pi} Im[g_{eff}^{0}(\vec{k}, \omega)]  \cr
N_{eff}^{(1)}(\omega) & = & \int \frac{k dk}{2 \pi} Im[g_{eff}^{1}(\vec{k}, \omega)]
\ea

As pointed out in the main paper, $N^{(0)/(1)}_{eff}(E_F) = \tfrac{N_0(E_F)}{2} (1 + O(\gbar))$ in the dilute limit; hence, we will approximate $N^{(0)/(1)}_{eff}(E_F) \approx \tfrac{N_0(E_F)}{2} = \tfrac{|E^2_F|}{4 \pi v_F^2}$ and $N^{(1)}_{eff}(E_F) \approx \tfrac{N_0(E_F)}{2} sgn(E_F)$. This finally gives the result for the quasi-particle lifetime near the Fermi surface, i.e. $\gbar = \gbar(k_F, E_F) = Im[\overline{\overline{\Sigma}}^{(R)}(k_F, E_F)]$, which is shown below. The real part of the self-energy that renormalizes $v_F$ and $\mu$ are ignored here, as $v_F$ and $\mu$ are taken to be experimentally determined parameters.
\begin{subequations}
\ba
\label{eqn: qp scattering rate}
\gbar & = & \gamma_0 \mI + \gamma_a \bl \cos \theta \, \sigma^x + \sin \theta \, \sigma^y \br \cr
& - & \, \gamma_b (\sin \theta \, \sigma^x - \cos \theta \, \sigma^y) + i \gamma_3 \, \sigma^z \\
\label{eqn: gamma_0}
 \gamma_0 & = & n_i N^{(0)}_{eff}(E_F) \left[ |T^0|^2 -  2\left( |T^3|^2 + |T^S|^2 - |T^A|^2 \right) \right]  \cr
  \label{eqn: gamma_a}
 \gamma_a & = & -4 n_i N^{(1)}_{eff}(E_F) \left[ |T^S|^2 + |T^A|^2 \right]   \cr
 \label{eqn: gamma_b}
 \gamma_b & = & 2 n_i N^{(0)}_{eff}(E_F) \left[ |T^0| |T^A| + |T^3| |T^S| \right]  \cr
 \label{eqn: gamma_3}
 \gamma_3 & = & 4 n_i N^{(1)}_{eff}(E_F)  \left[ |T^0| |T^S| + |T^3| |T^A| \right] 
\ea
\end{subequations}

The effective Greens function in the dilute impurity limit is now given by,
\begin{subequations}
\ba
\label{eqn: eff Greens func}
\Gbar^{(R)}_{eff}(\vec{k}, \omega) & = & \left[ \omega + \mu - v_F \vec{k} \cdot \vec{\sigma} - i \gbar(\vec{k}, \omega) \right]^{-1} \cr
& = & g^0_{eff}(k ,\omega) \mI + g^a_{eff}(k, \omega) \bl \cos \theta \, \sigma^x + \sin \theta \, \sigma^y \br \cr
 & + & g^b_{eff}(k, \omega) \bl \sin \theta \, \sigma^x - \cos \theta \, \sigma^y \br + g^3_{eff}(k, \omega) \sigma^z \\
  \label{eqn: G coeffs}
 g^0_{eff}(k, \omega) & = & \frac{(\Omega(k) + i \kappa(k)) (\omega + \mu - i \gamma_0 )}{\Omega^2(k) + \kappa^2(k)} \cr
 g^a_{eff}(k, \omega) & = & \frac{(\Omega(k) + i \kappa(k)) (v_F |\vec{k}| + i \gamma_a )}{\Omega^2(k) + \kappa^2(k)} \cr
 g^b_{eff}(k, \omega) & = & \frac{i \gamma_b (\Omega(k) + i \kappa(k)) }{\Omega^2(k) + \kappa^2(k)} \cr
 g^3_{eff}(k, \omega) & = & -\frac{\gamma_3 (\Omega(k) + i \kappa(k)) }{\Omega^2(k) + \kappa^2(k)}
\ea
\end{subequations}
where the denominator terms are $\Omega(k) = (\omega + \mu)^2 - v_F^2 |\vec{k}|^2 - \gamma_0^2 + \gamma_a^2 + \gamma_b^2 - \gamma_3^2$, $\kappa(k) = 2 \bl (\omega + \mu) \gamma_0 + v_F |\vec{k}| \gamma_a \br$. 

\section{SHE \& Rashba Edelstein Effect Correlation Functions}

Within the Kubo formalism, the longitudinal charge conductivity and spin-Hall conductivity, $\sigma_{yy}$ and $\sigma^{z}_{xy}$, are given by the retarded current-current and spin current-current correlation functions respectively,
\begin{subequations}
\ba
\pi^{(R)}_{yy}(\vec{k}, \omega) & = & -i \int_{- \infty}^{\infty} dt \, \me^{i \omega t} \theta(t) \, \langle [j_y(\vec{k}, t), j_y(\vec{k},0]) \rangle \\
\pi^{z,(R)}_{xy}(\vec{k}, \omega) & = & -i \int_{- \infty}^{\infty} dt \, \me^{i \omega t} \theta(t) \, \langle [\mathcal{J}^z_x(\vec{k}, t), j_y(\vec{k},0]) \rangle
\ea
\end{subequations}
Similarly, it is straightforward to derive a Kubo formula for the spin-accumulation due longitudinal charge transport, i.e. the Rashba-Edelstein effect.
\begin{subequations}
\ba
\langle \vec{S} \rangle & = & \Lim{\omega \rightarrow 0} \Lim{\vec{k} \rightarrow 0} \frac{E_{\alpha}}{\omega} \me^{i (\vec{k} \cdot \vec{r} - \omega t)} \int_{-\infty}^{\infty} d t' \theta(t') \langle [\vec{S}(\vec{k}, t'), j_{\alpha}(\vec{k}, 0)] \rangle \\
\pi^{i,(R)}_{\alpha}(\vec{k}, \omega) & = & -i \int_{-\infty}^{\infty} d t' \me^{i \omega t'}  \langle [S^i(\vec{k}, t'), j_{\alpha}(\vec{k}, 0)] \rangle
\ea
\end{subequations}

The spin current $\mathcal{J}^z_x$ has two components, one is the conventional spin current $j^z_x$ due to band-bending effects, and the other is the spin-torque current $P^z_x$, which are defined as follow,
\begin{subequations}
\ba
\label{eqn: spin current defns}
j^z_x(\vec{k}, \tau) & = & \sum_{\vec{k}_1} c\dg_{\vec{k}_1, \sigma}(\tau) \frac{(\vec{k} + \vec{k}_1)_x}{m} \sigma^z_{\sigma \sigma'} c_{\vec{k} + \vec{k}_1, \sigma'}(\tau) \\
P^z_x(\vec{k}, \tau) & = & \frac{2 i v_F}{k_x} \sum_{\vec{k}_1} c\dg_{\vec{k}_1, \sigma}(\tau) \Big( (\vec{k}_1 + \frac{\vec{k}}{2})_x \sigma^y - (\vec{k}_1 + \frac{\vec{k}}{2})_y \sigma^x\Big)_{\sigma \sigma'} c_{\vec{k} + \vec{k}_1, \sigma'}(\tau)
\ea
\end{subequations}
We will now separate the SHE into two contributions, $\pi^{z(1)}_{xy}$ and $\pi^{z(2)}_{xy}$, coming from the conventional spin current and the spin torque current respectively. All the Matsubara correlation functions, $\pi_{yy}(\vec{k}, i \omega_n)$, $\pi^{i}_{y}(\vec{k}, i \omega_n)$, $\pi^{z(1)}_{xy}(\vec{k}, i \omega_n)$ and $\pi^{z(2)}_{xy}(\vec{k}, i \omega_n)$, are given below, and analytic continuation ($i \omega_n \rightarrow \omega + i \eta$) will give the corresponding retarded correlation functions. 
\begin{subequations}
\label{eqn: Matsubara corr funcs}
\ba
\pi_{yy} (\vec{k}, i \omega_n) = - \int_0^{\beta} d \tau \me^{- i \omega_n \tau} \langle T_{\tau} U(\beta, 0) j_y(\vec{k}, \tau) j_y(\vec{k}, 0) \rangle \\
\pi^i_{y} (\vec{k}, i \omega_n) = - \int_0^{\beta} d \tau \me^{- i \omega_n \tau} \langle T_{\tau} U(\beta, 0) S^i(\vec{k}, \tau) j_y(\vec{k}, 0) \rangle \\
\pi^{z, (1)}_{xy} (\vec{k}, i \omega_n) = - \int_0^{\beta} d \tau \me^{- i \omega_n \tau} \langle T_{\tau} U(\beta, 0) j^z_x(\vec{k}, \tau) j_y(\vec{k}, 0) \rangle \\
\pi^{z, (2)}_{xy} (\vec{k}, i \omega_n) = - \int_0^{\beta} d \tau \me^{- i \omega_n \tau} \langle T_{\tau} U(\beta, 0) P^z_x(\vec{k}, \tau) j_y(\vec{k}, 0) \rangle
\ea
\end{subequations}

The correlation functions are written in the interaction representation, and $U(\beta, 0)$ is the S-matrix, which can be formally expanded as an infinite series of interacting terms involving $H^{int}$. Hence, the correlation functions are evaluated by expanding the S-matrix, and we show the expansion for $\pi^{z, (1)}_{xy}(\vec{k}, \tau)$ below.
\ba
\label{eqn: S matrix expansion}
\pi^{z,(1)}_{xy} (\vec{k}, \tau) & = & - \sum_{n=0}^{\infty} \frac{(-1)^n}{n!} \int_{0}^{\beta} d \tau_1 \ldots \int_{0}^{\beta} d \tau_n \langle T_{\tau} j^z_x(\vec{k}, \tau) H^{int}(\tau_1) \ldots H^{int}(\tau_n) j_y(\vec{k}, 0) \rangle
\ea

The $n = 0$ term in Eq.~\ref{eqn: S matrix expansion} is just the bare bubble diagram, and the $n = 2$ term will give the first correction to the scattering vertex.
\ba
\label{eqn: 1st order vertex correction}
\pi^{z, (1, n=2)}_{xy} (\vec{k}, i \omega_n) & = & - \int_{0}^{\beta} d \tau \int_{0}^{\beta} d \tau_1 \int_{0}^{\beta} d \tau_2 \me^{- i \omega_n \tau} \sum_{\vec{k}_1, \vec{k}_2} \frac{e v_F}{c} \langle T_{\tau} c\dg_{\vec{k}_1, \sigma}(\tau) \frac{(\vec{k} + \vec{k}_{1})_x}{m} \sigma^z_{\sigma \sigma'}  c_{\vec{k} + \vec{k}_1, \sigma'}(\tau) \cr
 & & \times H^{int}(\tau_1) H^{int}(\tau_2) c\dg_{\vec{k}_2, \nu}(0) \sigma^{y}_{\nu \nu'} c_{\vec{k} + \vec{k}_2, \nu'} \rangle \cr
& = & -\frac{e v_F}{m c} \sum_{\vec{p}, \vec{q}} \frac{1}{\beta} \sum_{i \omega_1} \sigma^z_{\sigma \sigma'} G_{\sigma' \mu_1}(\vec{p} + \vec{k}, i \omega_1 + i \omega_n) V_{\mu_1 \mu_2}(\vec{p} + \vec{k}, \vec{p} + \vec{q}) \cr
 & & \times G_{\mu_2 \nu}(\vec{p} + \vec{q}, i \omega_1 + i \omega_n) \sigma^{y}_{\nu \nu'} G_{\nu' \mu_3}(\vec{p} + \vec{q} - \vec{k}, i \omega_1) \cr
 & & \times V_{\mu_3 \mu_4}(\vec{p} + \vec{q} - \vec{k}, \vec{p}) G_{\mu_4 \sigma}(\vec{p}, i \omega_1) (\vec{k}_1 + \vec{k})_x
\ea
This corresponds to the Feynman diagram for the vertex correction from a single scattering event. Notice that only elastic scattering is considered here, as each scattering event does not change the energy of the electron; hence, all the Green's functions on the upper (and lower) legs of the bubble diagram have the same energy, e.g. in Eq.~\ref{eqn: 1st order vertex correction}, $G_{\sigma' \mu_1}(\vec{p} + \vec{k}, i \omega_1 + i \omega_n)$ and $G_{\mu_2 \nu}(\vec{p} + \vec{q}, i \omega_1 + i \omega_n)$ undergo a change of momentum and spin upon scattering off $V_{\mu_1 \mu_2}(\vec{p} + \vec{k}, \vec{p} + \vec{q})$, but do not exchange energy with the impurity.

Since energy is conserved in the upper and lower legs of the bubble diagram, we can now include the effect of all the scattering events from a single impurity on the vertex correction by replacing the scattering potential $V_{\mu_1 \mu_2}(\vec{k}, \vec{k}')$ by the full $\Tbar$-matrix to obtain,
\ba
\label{eqn: scattering vertex with T matrix}
\pi^{z, (1, T)}_{xy} (\vec{k}, i \omega_n) & = & -\frac{e v_F}{c} \sum_{\vec{p}, \vec{q}} \frac{1}{\beta} \sum_{i \omega_1}  \frac{(\vec{p} + \vec{k})_x}{m}
\Tr \Big[ \sigma^z \Gbar(\vec{p} + \vec{k}, i \omega_1 + i \omega_n) \Tbar(\vec{p} + \vec{k}, \vec{p} + \vec{q}) \cr
 & & \times \Gbar(\vec{p} + \vec{q}, i \omega_1 + i \omega_n) \sigma^{y} \Gbar(\vec{p} + \vec{q} - \vec{k}, i \omega_1) \Tbar(\vec{p} + \vec{q} - \vec{k}, \vec{p}) \Gbar(\vec{p}, i \omega_1) \Big] 
 \ea

Finally, scattering events from all the impurities can be included by defining a scattering vertex $\Gammabar^{y}(\vec{p} + \vec{k}, \vec{k}, i \omega_1 + i \omega_n, i \omega_n)$, whereby an infinite subset of scattering events are included in the Bethe-Salpeter equation,
\ba
\label{eqn: vertex Dyson eqn}
\Gammabar^{y}(\vec{p} + \vec{k}, \vec{k}, i \omega_1 + i \omega_n, i \omega_n) & = & \sigma^y + \sum_{\vec{q}} \Tbar(\vec{p} + \vec{k}, \vec{p} + \vec{q}, i \omega_1 + i \omega_n) \Gbar_{eff}(\vec{p}+\vec{q}, i \omega_1 + i \omega_n) \cr
& & \times \Gammabar^{y}(\vec{p} + \vec{q}, \vec{q}, i \omega_1 + i \omega_n, i \omega_n) \Gbar_{eff}(\vec{q}, i \omega_n) \Tbar(\vec{q}, \vec{k}, i \omega_n)
\ea
and the full correlation function is therefore,
\ba
\label{eqn: full scattering vertex with Gammabar}
\pi^{z, (1)}_{xy} (\vec{k}, i \omega_n) & = & -\frac{e v_F}{c} \sum_{\vec{p}} \frac{1}{\beta} \sum_{i \omega_1} \frac{(\vec{p} + \vec{k})_x}{m} \cr
 & & \times \Tr \Big[  \Gbar(\vec{p}, i \omega_1) \sigma^z  \Gbar(\vec{p} + \vec{k}, i \omega_1 + i \omega_n) \Gammabar^{y}(\vec{p} + \vec{k}, \vec{p}, i \omega_1 + i \omega_n, i \omega_1) \Big]  
 \ea

This infinite subset of ladder diagrams includes all the scattering corrections to the vertex from all the impurities, but does not include diagrams where scattering events from different impurities cross each other, i.e. this is the non-crossing approximation, which is reasonable in the dilute impurity limit.

Now let us evaluate the uniform limit of the Matsubara correlation function, $\underset{\vec{k} \to 0}{\lim} \, \pi^{z, (1)}_{xy}(\vec{k}, i \omega_n)$, by first doing the sum over the $i  \omega_1$ frequencies using the standard method of integrating over the poles of $n_F(z) = (\me^{\beta z} + 1)^{-1}$ in the complex $z$-plane. The poles of $n_F(z)$ are at $z = i \tfrac{2 \pi (n + 1)}{\beta}$, with residue of $- \tfrac{1}{\beta}$, and the sum $\sum_{i \omega_1}$ is replaced by an integration over the complex plane,
\ba
\pi^{z, (1)}_{xy}(\vec{k} = 0, i \omega_n) & = & - \frac{e v_F}{m c} \int \frac{d z}{2 \pi i} \mathcal{P}(z, z + i \omega_n) n_F(z) \cr
\mathcal{P}(z, z + i \omega_n) & =  & \sum_{\vec{p}} \vec{p}_x \Tr \Big[ \Gbar(\vec{p}, z) \sigma^z \Gbar(\vec{p}, z + i \omega_n) \Gammabar^{y}(\vec{p}, \vec{p}, z, z + i \omega_n) \Big]
\ea
The integral over the complex $z$-plane will also pick up the branch cuts of the Green's function $\Gbar(\vec{p}, z)$ and $\Gbar(\vec{p}, z + i \omega_n)$, which leads to branch cuts at $z = v_F |\vec{p}| - \mu = \xi(\vec{p})$ and $z + i \omega_n = v_F |\vec{p}| - \mu = \xi(\vec{p})$, and the upper ($\epsilon + i \delta$) and lower ($\epsilon - i \delta$)  paths along the branch cuts will give the following retarded and advanced contributions to the correlation function.

\ba
\pi^{z, (1)}_{xy}(\vec{k} = 0, i \omega_n) & = & - \int \frac{d \epsilon}{2 \pi i} n_F(\epsilon) \Big[ \mathcal{P}(\epsilon + i \delta, \epsilon + i \omega_n) - \mathcal{P}(\epsilon - i \delta, \epsilon + i \omega_n) \cr
 & & + \mathcal{P}(\epsilon - i \omega_n, \epsilon + i \delta) - \mathcal{P}(\epsilon - i \omega_n, \epsilon - i \delta) \Big]
\ea
Therefore, the retarded correlation function is obtained by analytic continuation $i \omega_n \rightarrow \omega + i \delta$,
\ba
\pi^{z, (1)}_{xy}(\vec{k} = 0, \omega) & = & -\frac{e v_F}{m c} \int \frac{d \epsilon}{2 \pi i} (n_F(\epsilon) - n_F(\epsilon + \omega))  \mathcal{P}(\epsilon - i \delta, \epsilon + \omega + i \delta) \cr
 & & - n_F(\epsilon) \mathcal{P}(\epsilon + i \delta, \epsilon + \omega + i \delta) + n_F(\epsilon + \omega) \mathcal{P}(\epsilon - i \delta, \epsilon + \omega - i \delta)
\ea
Following the standard discussion in \cite{Mahan2000}, the most singular contribution comes from $\mathcal{P}(\epsilon - i \delta, \epsilon + \omega + i \delta)$. Since the SHE conductivity is given by $\sigma^{z}_{xy}(\omega = 0) = - \underset{\omega \to 0}{\lim} Im [ \tfrac{\pi^z_{xy}(\vec{k} =0, \omega)}{\omega} ]$, hence we will calculate the following contribution to the retarded SHE correlation function.
\ba
\label{eqn: retarded SHE correlation func}
\pi^{z, (1)}_{xy}(\vec{k} = 0, \omega) & = & -\frac{e v_F}{m c} \int \frac{d \epsilon}{2 \pi i} (n_F(\epsilon) - n_F(\epsilon + \omega))  \mathcal{P}(\epsilon - i \delta, \epsilon + \omega + i \delta) \cr
\sigma^{z, (1)}_{xy}(\vec{k} = 0, \omega =0) & = & -Im \Big[\frac{e v_F}{m c} \int \frac{d \epsilon}{2 \pi i} \frac{d n_F(\epsilon)}{d \epsilon}  \mathcal{P}(\epsilon - i \delta, \epsilon + i \delta) \Big] \cr
\mathcal{P}(\epsilon - i \delta, \epsilon + i \delta) & =  & \sum_{\vec{p}} \vec{p}_x \Tr \Big[ \Gbar^{(A)}(\vec{p}, \epsilon) \sigma^z \Gbar^{(R)}(\vec{p}, \epsilon) \Gammabar^{y}(\vec{p}, \vec{p}, \epsilon - i \delta, \epsilon + i \delta) \Big]
\ea

The other correlation functions for the spin-torque current contribution to the SHE ($\pi^{z, (2)}_{xy}(\vec{k}, \omega)$), the Rashba-Edelstein effect ($\pi^{i}_{y}(\vec{k}, \omega)$), and the charge current conductivity ($\pi_{yy}(\vec{k}, \omega)$) are derived in a similar manner, and we obtain,
\ba
\label{eqn: pizxy corr function}
\pi_{yy}(\vec{k} = 0, \omega) & = & \underset{\vec{k} \to 0}{\lim} \bl \frac{e v_F}{c} \br^2 \int_{-\infty}^{\infty} \frac{d \epsilon}{2 \pi i} \int \frac{d^2 p}{(2 \pi)^2} \bl n_F(\epsilon) - n_F(\epsilon + \omega) \br   \cr
 & & \times \Tr \left[ \Gbar^{(A)}(\vec{p}, \epsilon) \sigma^y \Gbar^{(R)}(\vec{p} + \vec{k}, \epsilon) \Gammabar^{y}(\vec{p}, \vec{k}_1 + \vec{k}, \epsilon) \right] \\
\pi^{z, (2)}_{xy}(\vec{k} = 0, \omega) & = & \pi^{z, (2a)}_{xy}(\vec{k} = 0, \omega) + \pi^{z, (2b)}_{xy}(\vec{k} = 0, \omega) \\
\pi^{z, (2a)}_{xy}(\vec{k} = 0, \omega) & = & \underset{\vec{k} \to 0}{\lim} \frac{2 i e v_F^2}{c} \int_{-\infty}^{\infty} \frac{d \epsilon}{2 \pi i} \int \frac{d^2 p}{(2 \pi)^2} \frac{p_{y} + \frac{k_y}{2}}{k_x} \bl n_F(\epsilon) - n_F(\epsilon + \omega) \br \cr
& & \times \Tr \left[ \Gbar^{(A)}(\vec{p}, \epsilon) \sigma^x \Gbar^{(R)}(\vec{p} + \vec{k}, \epsilon) \Gammabar^{y}(\vec{p}, \vec{p} + \vec{k}, \epsilon) \right] \cr
\pi^{z, (2b)}_{xy}(\vec{k} = 0, \omega) & = & \underset{\vec{k} \to 0}{\lim} \frac{2 i e v_F^2}{c} \int_{-\infty}^{\infty} \frac{d \epsilon}{2 \pi i} \int \frac{d^2 p}{(2 \pi)^2} \frac{p_x + \frac{k_x}{2}}{k_x} \bl n_F(\epsilon + \omega) - n_F(\epsilon) \br \cr
& & \times \Tr \left[ \Gbar^{(A)}(\vec{p}, \epsilon) \sigma^y \Gbar^{(R)}(\vec{p} + \vec{k}, \epsilon) \Gammabar^{y}(\vec{p}, \vec{p} + \vec{k}, \epsilon) \right] \cr
\pi^{i}_{y}(\vec{k} = 0, \omega) & = & \underset{\vec{k} \to 0}{\lim} \frac{e v_F}{c} \int_{-\infty}^{\infty} \frac{d \epsilon}{2 \pi i} \int \frac{d^2 p}{(2 \pi)^2} \bl n_F(\epsilon + \omega) - n_F(\epsilon) \br \cr
& & \times \Tr \left[ \Gbar^{(A)}(\vec{p}, \epsilon) \sigma^i \Gbar^{(R)}(\vec{p} + \vec{k}, \epsilon) \Gammabar^{y}(\vec{p}, \vec{p} + \vec{k}, \epsilon) \right] 
\ea

\section{Vertex Correction}

For four fermion correlation functions, like the current-current and spin current-current correlation functions, we have to consider the effects of impurity scattering on the scattering vertex\cite{Mahan2000}, in addition to the quasi-particle self-energy corrections. This arises from an infinite subset of Feynman ladder diagrams shown in the main paper, and is summed up in the Bethe Salpeter equation for the scattering vertex $\Gammabar^{y}(\vec{k} + \vec{p}, \vec{p}, i \omega_1 + i \omega_n, i \omega_n)$ (Eq.~\ref{eqn: vertex Dyson eqn}).
\ba
\label{eqn: vertex Dyson eqn}
\Gammabar^{y}(\vec{k} + \vec{p}, \vec{p}, i \omega_1 + i \omega_n, i \omega_n) & = & \sigma^y + \sum_{\vec{q}} \Tbar(\vec{k} + \vec{p}, \vec{k} + \vec{q}, i \omega_1 + i \omega_n) \Gbar_{eff}(\vec{k}+\vec{q}, i \omega_1 + i \omega_n) \cr
& & \times \Gammabar^{y}(\vec{k} + \vec{q}, \vec{q}, i \omega_1 + i \omega_n, i \omega_n) \Gbar_{eff}(\vec{q}, i \omega_n) \Tbar(\vec{q}, \vec{p}, i \omega_n)
\ea
Here, $\vec{k}$ and $i \omega_1$ are the  external momentum and frequency, and the DC uniform limit of the conductivities are obtained by analytic continuation of $i \omega_1 \rightarrow \omega + i \eta$, setting the limit $\vec{k} \rightarrow 0$, and then setting $\omega \rightarrow 0$, i.e. $\underset{\omega \to 0}{\lim} \; \underset{\vec{k} \to 0}{\lim} $. Hence, we only need to calculate the on-shell component of the scattering vertex $\Gammabar^{y}(\vec{p}, \omega) = \Gammabar^{y}(\vec{p}, \omega - i \eta, \omega + i \eta)$, which is defined by,
\ba
\Gammabar^{y}(\vec{p}, \omega) & =  & \sigma^y + \sum_{\vec{q}} \Tbar(\vec{p}, \vec{q}, \omega + i \eta) \Gbar_{eff}(\vec{q}, \omega + i \eta) \cr
& & \times \Gammabar^{y}(\vec{q}, \omega) \Gbar_{eff}(\vec{q}, \omega - i \eta) \Tbar(\vec{q}, \vec{p}, \omega - i \eta) \cr
 & =  & \sigma^y + \sum_{\vec{q}}  \Tbar^{(R)}(\vec{p}, \vec{q}, \omega) \Gbar_{eff}^{(R)}(\vec{q}, \omega) \Gammabar^{y}(\vec{q}, \omega) \Gbar_{eff}^{(A)}(\vec{q}, \omega) \Tbar^{(A)}(\vec{q}, \vec{p}, \omega)
\ea

Note that both the advanced and retarded Green's function and $\Tbar$-matrices, $\Gbar_{eff}^{(R)}(\vec{p}, \omega)$, $\Gbar_{eff}^{(A)}(\vec{p}, \omega)$, $\Tbar^{(R)}(\vec{p}, \vec{q}, \omega)$ and $\Tbar^{(A)}(\vec{p}, \vec{q}, \omega)$ enter into the Bethe-Salpeter equation due to the branch cut in the complex plane, when the integral over the complex plane is carried out. Similar to the assumption for the $\Tbar$-matrix, the scattering vertex is assumed to be momentum-independent near $E_F$, and we will do a similar multipole expansion of $\Gammabar^{y}(|\vec{p}| = k_F, \theta, \omega = E_F) = \sum_{n} \Gamma^i_n \me^{i n \theta} \sigma^i$, keeping only the $l = 0$ and $l = 1$ scattering channels. 
\be
\Gammabar^{y}(|\vec{p}| = k_F, \theta, \omega = E_F) =  \Gamma^i_0 \sigma^i + \left[ \Gamma^0_{p_x} \cos \theta + i \Gamma^0_{p_y} \sin \theta \right] \mI + \left[ \Gamma^i_{p_x} \cos \theta + i \Gamma^i_{p_y} \sin \theta \right] \sigma^i 
\ee
Hence, the Bethe-Salpeter equation is reduced to,
\ba
\Gammabar^{y}(\vec{p}, \omega) & = & \sigma^y + \int \frac{d \theta_q}{2 \pi} \Tbar^{(R)}(|\vec{p}| = |\vec{p} + \vec{q}| = k_F, \theta_p, \theta_{p + q}, \omega) \\
 & & \times \Bigg[ \int \frac{q dq}{2 \pi} \Gbar^{(R)}(\vec{p} + \vec{q}, \omega) \Gammabar^{y}(\vec{p} + \vec{q}, \omega) \Gbar^{(A)}(\vec{p} + \vec{q}, \omega) \Bigg] \cr
 & & \times \Tbar^{(A)}(|\vec{p} + \vec{q}| = |\vec{q}| = k_F, \theta_{p + q}, \theta_q, \omega) \cr
 \sum_{n} \Gamma^i_n \me^{i n \theta} \sigma^i & = & \sigma^y + \sum_{n_1 ... n_7} \int \frac{d \theta_q}{2 \pi} T^{i_1}_{n_1 n_2} \me^{i (n_1 \theta_k - n_2 \theta_{k+q})} T^{i_5}_{n_6 n_7} \me^{i (n_6 \theta_k - n_7 \theta_{k+q})} \sigma^{i_1} \sigma^{i_2} \sigma^{i_3} \sigma^{i_4} \sigma^{i_5}  \cr
 & & \times \Bigg[ \int \frac{q dq}{2 \pi} g_{n_3}^{i_2, (R)}(|\vec{p} + \vec{q}|, \omega) \me^{- i n_3 \theta_{p+q}}  \Gamma^{i_3}_{n_4} \me^{-i n_4 \theta_{p+q}}  g_{n_5}^{i_4, (A)}(|\vec{p} + \vec{q}|, \omega) \me^{- i n_5 \theta_{p+q}} \Bigg] \nonumber
\ea 

Since the $\Gamma^i_n$ coefficients are assumed to be invariant near $k_F$, the $\int dq$-integral is carried out over all the spin and angular momentum resolved Green's function components, $g_{m}^{i, (R)}(|\vec{p} + \vec{q}|, \omega) \, g_{n}^{j, (A)}(|\vec{p} + \vec{q}|, \omega)$. As the Weyl fermions are spin-momentum locked; hence, the spin $i$ and momentum $m$ indices are related, i.e. $m = 0$ for $i = [0,3]$, and $m = \pm 1$ for $i \in [1,2]$. We can now define, 
\be
\label{eqn: spectral weight integral}
\xi^{ij}(\epsilon) = \int \frac{k dk}{2 \pi} g^{i, (R)}(|\vec{k}|, \epsilon) g^{j, (A)}(|\vec{k}|, \epsilon)
\ee
We have carried out a change of variable from $\omega + \mu \rightarrow \epsilon$ here, thereby absorbing the factors of $\mu$ that appear in the Green's function into $\epsilon$, which is now the energy measured from $E_F$.

Knowing that $G^{(R)}(\vec{k}, \epsilon) G^{(A)}(\vec{k}, \epsilon) = \tfrac{A(\vec{k}, \epsilon)}{Im[\Sigma(\vec{k}, \epsilon)]} \approx \tfrac{A(\vec{k}, \epsilon)}{\gamma}$, this means that $\xi^{ij}(\epsilon)$ is basically the spin-resolved density of states divided by the quasi-particle scattering rate. The dominant terms are the $s$-wave, $p$-wave and $s-p$ spin-flip DOS, $\xi^{00}(\epsilon)$, $\xi^{aa}(\epsilon)$ and $\xi^{0a}(\epsilon) = (\xi^{a0}(\epsilon))^*$ respectively, which are calculated to be,
\ba
\xi^{00}(\epsilon) & = & \frac{1}{2 \pi v_F^2} \Big( \frac{\pi \epsilon}{2 (\gamma_0 + \gamma_a)} + \frac{\pi \gamma_0^2}{2(\gamma_0 + \gamma_a) \epsilon} \Big) \cr
\xi^{aa}(\epsilon) & = & \frac{1}{2 \pi v_F^2} \Big( \frac{\pi \epsilon}{2 (\gamma_0 + \gamma_a)} + \frac{\pi \gamma_a^2}{2(\gamma_0 + \gamma_a) \epsilon} \Big) \cr
\xi^{0a}(\epsilon) & = & \frac{1}{2 \pi v_F^2} \frac{\pi}{2 (\gamma_0 + \gamma_a)} (\epsilon - i \gamma_0) (1 - \frac{i \gamma_a}{\epsilon})
\ea

The above set of coupled equations for the $\Gammabar$-coefficients are then solved analytically, and the finite terms are shown below; and the other terms $\Gamma^0_0$, $\Gamma^3_0$, $\Gamma^1_{p_x}$, $\Gamma^1_{p_y}$, $\Gamma^2_{p_x}$ and $\Gamma^2_{p_y}$ are equal to zero. 
\begin{subequations}
\ba
\label{eqn: Vertex coeffs}
\Gamma^1_0(E_F) & = &  \Big[ 2 n_i (|T^3_1| |T^A| + |T^0| |T^3_0| - 2 i |T^A| |T^S|) (\xi^{00} + \xi^{aa}) \Big] \cr
 & &  \times \Bigg[ 1 - n_i \Big( |T^0|^2 + |T^3_0|^2 + 2 |T^S|^2 - 2 |T^A|^2 - 2 |T^3_1|^2 \Big) (\xi^{00} + \xi^{aa}) \cr
 &  & - 2 n_i \Big(|T^3_1|^2 - |T^3_0|^2 \Big) (\xi^{0a} + \xi^{a0}) \Bigg]^{-1} \cr
 & = & \frac{\gamma_{asym,1} + \gamma_{30}- i \gamma_{asym,3}}{\gamma_t} + O(\frac{\gamma}{E_F}) \\
 \cr
\Gamma^2_0(E_F) & = & \Bigg[ 1 - n_i \Big( |T^0|^2 + |T^3_0|^2 - 2 |T^3_1|^2 \Big) \xi^{00} - n_i \Big( |T^S|^2 + |T^A|^2\Big) \xi^{aa} + i \, n_i |T^3_1| |T^S| (\xi^{0a} + \xi^{a0}) \Bigg] \cr
 &  & \times \Bigg[ 1 - n_i \Big( |T^0|^2 + |T^3_0|^2 + 2 |T^S|^2 - 2 |T^A|^2 - 2 |T^3_1|^2 \Big) (\xi^{00} + \xi^{aa}) \cr
 &  & - 2 n_i \Big(|T^3_1|^2 - |T^3_0|^2 \Big) (\xi^{0a} + \xi^{a0}) \Bigg]^{-1} \cr
 & = & \frac{\gamma_0 + \gamma_a}{\gamma_t} + i \frac{\gamma_{31,s}}{\gamma_t} + O(\frac{\gamma}{E_F}) \\
 \cr
\Gamma^0_{p_x}(E_F) & = & n_i \Big[ \Big( |T^A|^2| + |T^S|^2 \Big) (\xi^{0a} + \xi^{a0}) + 2 |T^0| |T^3_1| (\xi^{0a} - \xi^{a0}) \Big] \cr
& & \times \Big[ 2 n_i (|T^3_1| |T^A| + |T^0| |T^3_0| - 2 i |T^A| |T^S|) (\xi^{00} + \xi^{aa}) \Big] \cr
 & &  \times \Bigg[ 1 - n_i \Big( |T^0|^2 + |T^3_0|^2 + 2 |T^S|^2 - 2 |T^A|^2 - 2 |T^3_1|^2 \Big) (\xi^{00} + \xi^{aa}) \cr
 &  & - 2 n_i \Big(|T^3_1|^2 - |T^3_0|^2 \Big) (\xi^{0a} + \xi^{a0}) \Bigg]^{-1} \cr
 & & + n_i \Big[ 2 |T^A| |T^3_1| (\xi^{00} + \xi^{aa}) - 2 i |T^A| |T^S| (\xi^{0a} + \xi^{a0}) \Big] \cr
 & & \times \Bigg[ 1 - n_i \Big( |T^0|^2 + |T^3_0|^2 - 2 |T^3_1|^2 \Big) \xi^{00} - n_i \Big( |T^S|^2 + |T^A|^2\Big) \xi^{aa} + i \, n_i |T^3_1| |T^S| (\xi^{0a} + \xi^{a0}) \Bigg] \cr
 &  & \times \Bigg[ 1 - n_i \Big( |T^0|^2 + |T^3_0|^2 + 2 |T^S|^2 - 2 |T^A|^2 - 2 |T^3_1|^2 \Big) (\xi^{00} + \xi^{aa}) \cr
 &  & - 2 n_i \Big(|T^3_1|^2 - |T^3_0|^2 \Big) (\xi^{0a} + \xi^{a0}) \Bigg]^{-1} \cr
& = & \frac{\gamma_{asym,1} - i \gamma_{asym,3}}{\gamma_t}  - \frac{\gamma_a (\gamma_{30} + \gamma_{asym,1} - i \gamma_{asym,3})}{4 \gamma_t (\gamma_0 + \gamma_a)} + O(\frac{\gamma}{E_F}) \\
 \cr
 \Gamma^0_{p_y}(E_F) & = & 2 i \, n_i \Big[ |T^3_1| |T^A| (\xi^{00} + \xi^{aa}) + |T^A| |T^S| (\xi^{0a} + \xi^{a0}) \Big] \cr
 & & \times \Big[ 2 n_i (|T^3_1| |T^A| + |T^0| |T^3_0| - 2 i |T^A| |T^S|) (\xi^{00} + \xi^{aa}) \Big] \cr
 & &  \times \Bigg[ 1 - n_i \Big( |T^0|^2 + |T^3_0|^2 + 2 |T^S|^2 - 2 |T^A|^2 - 2 |T^3_1|^2 \Big) (\xi^{00} + \xi^{aa}) \cr
 &  & - 2 n_i \Big(|T^3_1|^2 - |T^3_0|^2 \Big) (\xi^{0a} + \xi^{a0}) \Bigg]^{-1} \cr
 & & - i \, n_i \Big[ |T^S|^2 + |T^A|^2 \Big] (\xi^{0a} + \xi^{a0}) \cr
 & & \times \Bigg[ 1 - n_i \Big( |T^0|^2 + |T^3_0|^2 - 2 |T^3_1|^2 \Big) \xi^{00} - n_i \Big( |T^S|^2 + |T^A|^2\Big) \xi^{aa} + i \, n_i |T^3_1| |T^S| (\xi^{0a} + \xi^{a0}) \Bigg] \cr
 &  & \times \Bigg[ 1 - n_i \Big( |T^0|^2 + |T^3_0|^2 + 2 |T^S|^2 - 2 |T^A|^2 - 2 |T^3_1|^2 \Big) (\xi^{00} + \xi^{aa}) \cr
 &  & - 2 n_i \Big(|T^3_1|^2 - |T^3_0|^2 \Big) (\xi^{0a} + \xi^{a0}) \Bigg]^{-1} \cr
 & = & \frac{i}{4} \frac{\gamma_a}{\gamma_t} + \frac{i}{2} \frac{(\gamma_{asym,1} - i \gamma_{asym,3})(\gamma_{asym,1} - i \gamma_{asym,3} + \gamma_{30})}{\gamma_t(\gamma_0 + \gamma_a)} + O(\frac{\gamma}{E_F}) \\
 \cr
 \label{eqn: Gamma3px}
\Gamma^3_{p_x}(E_F) & = & 2 n_i \Big[ |T^3_0| |T^S| - i |T^0| |T^A| \Big] \xi^{00} \cr
& & \times  \Big[ 2 n_i (|T^3_1| |T^A| + |T^0| |T^3_0| - 2 i |T^A| |T^S|) (\xi^{00} + \xi^{aa}) \Big] \cr
 & &  \times \Bigg[ 1 - n_i \Big( |T^0|^2 + |T^3_0|^2 + 2 |T^S|^2 - 2 |T^A|^2 - 2 |T^3_1|^2 \Big) (\xi^{00} + \xi^{aa}) \cr
 &  & - 2 n_i \Big(|T^3_1|^2 - |T^3_0|^2 \Big) (\xi^{0a} + \xi^{a0}) \Bigg]^{-1} \cr
& & - 2 n_i \Big[ (|T^0| |T^S| + i |T^A| |T^3_0|) \xi^{00} - i |T^0| |T^3_1| (\xi^{0a} + \xi^{a0}) \Big] \cr
& & \times \Bigg[ 1 - n_i \Big( |T^0|^2 + |T^3_0|^2 - 2 |T^3_1|^2 \Big) \xi^{00} - n_i \Big( |T^S|^2 + |T^A|^2\Big) \xi^{aa} + i \, n_i |T^3_1| |T^S| (\xi^{0a} + \xi^{a0}) \Bigg] \cr
 &  & \times \Bigg[ 1 - n_i \Big( |T^0|^2 + |T^3_0|^2 + 2 |T^S|^2 - 2 |T^A|^2 - 2 |T^3_1|^2 \Big) (\xi^{00} + \xi^{aa}) \cr
 &  & - 2 n_i \Big(|T^3_1|^2 - |T^3_0|^2 \Big) (\xi^{0a} + \xi^{a0}) \Bigg]^{-1} \cr
& = & - \frac{\gamma_{s}}{\gamma_t} - i \frac{\gamma_{31} + \gamma_{asym,2}}{\gamma_t} + \frac{\gamma_{3s} \gamma_{asym,1}}{2 \gamma_t (\gamma_0 + \gamma_a)} \\
\cr
 \Gamma^3_{p_y}(E_F) & = & n_i \Big[ |T^0| |T^A| + i |T^3_0| |T^S| \Big] \xi^{00} \cr
 & & \times \Bigg[ 1 - n_i \Big( |T^0|^2 + |T^3_0|^2 - 2 |T^3_1|^2 \Big) \xi^{00} - n_i \Big( |T^S|^2 + |T^A|^2\Big) \xi^{aa} + i \, n_i |T^3_1| |T^S| (\xi^{0a} + \xi^{a0}) \Bigg] \cr
 &  & \times \Bigg[ 1 - n_i \Big( |T^0|^2 + |T^3_0|^2 + 2 |T^S|^2 - 2 |T^A|^2 - 2 |T^3_1|^2 \Big) (\xi^{00} + \xi^{aa}) \cr
 &  & - 2 n_i \Big(|T^3_1|^2 - |T^3_0|^2 \Big) (\xi^{0a} + \xi^{a0}) \Bigg]^{-1} \cr
& & - n_i \Big[ (|T^3_0| |T^A| + i |T^0| |T^S|) \xi^{00}  + |T^0| |T^3_1| (\xi^{0a} + \xi^{a0}) \Big] \cr
& & \times \Big[ 2 n_i (|T^3_1| |T^A| + |T^0| |T^3_0| - 2 i |T^A| |T^S|) (\xi^{00} + \xi^{aa}) \Big] \cr
 & &  \times \Bigg[ 1 - n_i \Big( |T^0|^2 + |T^3_0|^2 + 2 |T^S|^2 - 2 |T^A|^2 - 2 |T^3_1|^2 \Big) (\xi^{00} + \xi^{aa}) \cr
 &  & - 2 n_i \Big(|T^3_1|^2 - |T^3_0|^2 \Big) (\xi^{0a} + \xi^{a0}) \Bigg]^{-1} \cr
& = & -\frac{\gamma_3}{4 \gamma_t} - i \frac{\gamma_{3s}}{2 \gamma_t} + \frac{1}{2} \frac{(\gamma_{30} + \gamma_{asym,1})(\gamma_{31} + \gamma_{asym,2}) - \gamma_{s} \gamma_{asym,3}}{\gamma_t (\gamma_0 + \gamma_a)} \cr
& & - \frac{i}{2} \frac{\gamma_{s} (\gamma_{30} + \gamma_{asym,1}) + \gamma_{asym,3} (\gamma_{31} + \gamma_{asym,2})}{\gamma_t (\gamma_0 + \gamma_a)}
 \ea
\end{subequations}
Hence, using the results of $\xi^{ij}(E_F)$ listed above, the scattering vertex is,
\ba
\Gammabar^{y}(|\vec{k}| = k_F, \theta, E_F) & = & \Gamma^1_0(E_F) \mI + \Gamma^2_0(E_F) \, \sigma^y  + ( \Gamma^0_{p_x}(E_F) \mI + \Gamma^3_{p_x}(E_F) \, \sigma^z) \cos \theta \cr
& & + i \bl \Gamma^0_{p_y}(E_F) \mI + \Gamma^3_{p_y}(E_F) \, \sigma^z \br \sin \theta 
\ea

Since $\Gamma^2_0$ is the scattering vertex channel for longitudinal electrical conductivity, we have defined a transport scattering rate $\gamma_t = ( \tfrac{1}{2} \gamma_0 + \gamma_a - 2 \gamma_{t'})$, in terms of $\gamma_0$, $\gamma_a$, and an additional transport contribution, $\gamma_{t'} = 2 n_i \pi N_0(E_F) (|T^3_1|^2 - |T^3_0|^2)$. Since $\gamma_{t'} \propto V_0^4 V_1^2 N_0(E_F)^5$, it is much weaker than $\gamma_0 \propto V_0^2 N_0(E_F)$ and $\gamma_a \propto V_0^2 V_1^2 N_0(E_F)^3$, and we do not display $\gamma_{t'}$ in the main paper, but instead, display it here for completeness.

In addition, there are spin flip scattering rates arising from $|T^A|$ and $|T^S|$, $\gamma_{s} = \frac{n_i \pi N_0(E_F)}{2} |T^0| |T^S|$, $\gamma_{asym,1} = 2 n_i \pi N_0(E_F) |T^3_1| |T^A|$, $\gamma_{asym,2} = \frac{n_i \pi N_0(E_F)}{2} |T^3_0| |T^A|$, $\gamma_{asym,3} = \frac{n_i \pi N_0(E_F)}{2} |T^S| |T^A|$, $\gamma_{30} = \frac{n_i \pi N_0(E_F)}{2} |T^3_0| |T^0|$, $\gamma_{31} = \frac{n_i \pi N_0(E_F)}{2} |T^3_1| |T^0|$, $\gamma_{3s} = \frac{n_i \pi N_0(E_F)}{2} |T^3_0| |T^S|$ and $\gamma_{31,s} = \frac{n_i \pi N_0(E_F)}{2} |T^3_1| |T^S|$, which are proportional to $T^S$ and $T^A$, the symmetric and asymmetric component of the $\Tbar$-matrix, as well as the $\sigma^z$ components of the $\Tbar$-matrix, $T^3_0$ and $T^3_1$.

\section{Longitudinal Charge Transport and SHE DC Conductivities}

We calculate the longitudinal charge conductivity, the Rashba-Edelstein effect, and the spin torque contribution to the SHE here. The retarded correlation functions for the spin-torque current contribution to the SHE ($\pi^{z, (2)}_{xy}(\vec{k}, \omega)$), the Rashba-Edelstein effect ($\pi^{i}_{y}(\vec{k}, \omega)$), and the charge current conductivity ($\pi_{yy}(\vec{k}, \omega)$) are shown below, and the DC conductivities are all given by first taking the limit of $\lim \vec{k} \to 0$, then taking the DC limit of $\lim \omega \to 0$, $\sigma^{(\text{DC})} = - \underset{\omega \to 0}{\lim} \underset{\vec{k} \to 0}{\lim} \, Im [\tfrac{\pi(\vec{k}, \omega)}{\omega}]$.
\ba
\label{eqn: pizxy corr function}
\pi_{yy}(\vec{k} = 0, \omega) & = & \underset{\vec{k} \to 0}{\lim} \bl \frac{e v_F}{c} \br^2 \int_{-\infty}^{\infty} \frac{d \epsilon}{2 \pi i} \int \frac{d^2 p}{(2 \pi)^2} \Tr \left[ \Gbar^{(A)}(\vec{p}, \epsilon) \sigma^y \Gbar^{(R)}(\vec{p} + \vec{k}, \epsilon) \Gammabar^{y}(\vec{p}, \vec{p} + \vec{k}, \epsilon) \right] \cr
 & \times & \bl n_F(\epsilon) - n_F(\epsilon + \omega) \br \\
\pi^{z, (2)}_{xy}(\vec{k} = 0, \omega) & = & \pi^{z, (2a)}_{xy}(\vec{k} = 0, \omega) + \pi^{z, (2b)}_{xy}(\vec{k} = 0, \omega) \\
\pi^{z, (2a)}_{xy}(\vec{k} = 0, \omega) & = & \underset{\vec{k} \to 0}{\lim} \frac{2 i e v_F^2}{c} \int_{-\infty}^{\infty} \frac{d \epsilon}{2 \pi i} \int \frac{d^2 p}{(2 \pi)^2} \Tr \left[ \Gbar^{(A)}(\vec{p}, \epsilon) \sigma^x \Gbar^{(R)}(\vec{p} + \vec{k}, \epsilon) \Gammabar^{y}(\vec{p}, \vec{p} + \vec{k}, \epsilon) \right] \cr
 & \times & \frac{p_y + \frac{k_y}{2}}{p_x} \bl n_F(\epsilon) - n_F(\epsilon + \omega) \br \cr
\pi^{z, (2b)}_{xy}(\vec{k} = 0, \omega) & = & - \underset{\vec{k} \to 0}{\lim} \frac{2 i e v_F^2}{c} \int_{-\infty}^{\infty} \frac{d \epsilon}{2 \pi i} \int \frac{d^2 p}{(2 \pi)^2} \Tr \left[ \Gbar^{(A)}(\vec{p}, \epsilon) \sigma^y \Gbar^{(R)}(\vec{p} + \vec{k}, \epsilon) \Gammabar^{y}(\vec{p}, \vec{p} + \vec{k}, \epsilon) \right] \cr
& \times & \frac{p_x + \frac{k_x}{2}}{p_x} \bl n_F(\epsilon) - n_F(\epsilon  + \omega) \br \cr
\pi^{i}_{y}(\vec{k} = 0, \omega) & = & \underset{\vec{k} \to 0}{\lim} \frac{e v_F}{c} \int_{-\infty}^{\infty} \frac{d \epsilon}{2 \pi i} \int \frac{d^2 p}{(2 \pi)^2} \Tr \left[ \Gbar^{(A)}(\vec{p}, \epsilon) \sigma^i \Gbar^{(R)}(\vec{p} + \vec{k}, \epsilon) \Gammabar^{y}(\vec{p}, \vec{p} + \vec{k}, \epsilon) \right] \cr
& \times & \bl n_F(\epsilon + \omega) - n_F(\epsilon) \br
\ea
We have specialized to the case of a charge current along $\hat{y}$ in the expression for the Rashba-Edelstein effect. For the SHE Kubo formula, we have to Taylor expand the Green's function $\Gbar^{(R)}(\vec{p} + \vec{k}, \epsilon) = \Gbar^{(R)}(\vec{p}, \epsilon) + k_i \tfrac{d \Gbar^{(R)}(\vec{p}, \epsilon)}{d p_i}$, which is shown in detail below.

\begin{subequations}
\ba
\label{eqn: Greens func derivative}
\frac{d \Gbar^{(R)}(\vec{p}, \epsilon)}{d p_x} & = & \frac{\partial \Gbar^{(R)}(\vec{p}, \epsilon) }{\partial p} \frac{\partial p}{\partial p_x} + \frac{\partial \Gbar^{(R)}(\vec{p}, \epsilon) }{\partial \theta} \frac{\partial \theta}{\partial p_x} \\
\frac{\partial \Gbar^{(R)}(\vec{p}, \epsilon) }{\partial p} \frac{\partial p}{\partial p_x} & = &  \left[ \frac{d g^0}{d p} \mI + \frac{d g^3}{d p}  \sigma^z + \frac{d g^a}{d p} (\cos \theta_p \sigma^x + \sin \theta_p \sigma^y) + \frac{d g^b}{d p} (\sin \theta_p \sigma^x - \cos \theta_p \sigma^y)\right] \cos \theta_p  \cr
\frac{\partial \Gbar^{(R)}(\vec{p}, \epsilon) }{\partial \theta} \frac{\partial \theta}{\partial p_x} & = &  \left[ g^a (-\sin \theta_p \sigma^x + \cos \theta_p \sigma^y) + g^b (\cos \theta_p \sigma^x + \sin \theta_p \sigma^y) \right] \big( - \frac{\sin \theta_p}{p} \big) \cr
\frac{d \Gbar^{(R)}(\vec{p}, \epsilon)}{d p_y} & = & \frac{\partial \Gbar^{(R)}(\vec{p}, \epsilon) }{\partial p} \frac{\partial p}{\partial p_y} + \frac{\partial \Gbar^{(R)}(\vec{p}, \epsilon) }{\partial \theta} \frac{\partial \theta}{\partial p_y} \\
\frac{\partial \Gbar^{(R)}(\vec{p}, \epsilon) }{\partial p} \frac{\partial p}{\partial p_y} & = &  \left[ \frac{d g^0}{d p} \mI + \frac{d g^3}{d p}  \sigma^z + \frac{d g^a}{d p} (\cos \theta_p \sigma^x + \sin \theta_p \sigma^y) + \frac{d g^b}{d p} (\sin \theta_p \sigma^x - \cos \theta_p \sigma^y)\right] \sin \theta_p \cr
\frac{\partial \Gbar^{(R)}(\vec{p}, \epsilon) }{\partial \theta} \frac{\partial \theta}{\partial p_y} & = &  \left[ g^a (-\sin \theta_p \sigma^x + \cos \theta_p \sigma^y) + g^b (\cos \theta_p \sigma^x + \sin \theta_p \sigma^y) \right] \big( \frac{\cos \theta_p}{p} \big) \nonumber
\ea
\end{subequations}
Following the same approximation of an average $\Gammabar$-matrix near $E_F$, the spin current-current correlation function is then given in terms of the $\Gammabar$-coefficients, and the spin-resolved density of states $\xi^{ij}(E_F)$, as well as the quantity involving the integral of $\Gbar^{(A)}(\vec{k}, \epsilon) \tfrac{d \Gbar^{(R)}(\vec{k}, \epsilon)}{d \vec{k}}$, which we term $\eta^{ij} (\epsilon)$,
\begin{subequations}
\ba
\label{eqn: Green func integral}
\eta^{ij}(\epsilon) & \equiv & \int_{-\infty}^{\infty} \frac{d p}{2 \pi} \, p^2 \, \frac{d g^{i,(R)}(p, \epsilon)}{d p} g^{j,(A)}_{eff}(p, \epsilon) \\
\cr
\eta^{00}(\epsilon) & = & \int \frac{d p}{2 \pi} v_F p^2 \Bigg[ \frac{2(- v_F p + i \gamma_a) (\epsilon - i \gamma_0) }{ \Omega(p)^2 + \kappa(p)^2} \cr
 & & + \frac{4 (v _F p \Omega(p) - \gamma_a \kappa(p)) (\Omega(p) + i \kappa(p)) (\epsilon - i \gamma_0) }{ (\Omega(p)^2 + \kappa(p)^2)^2 }  \Bigg] \frac{ (\Omega(p) - i \kappa(p)) (\epsilon + i \gamma_0) }{ \Omega(p)^2 + \kappa(p)^2} \cr
  & = & \frac{1}{2 \pi v_F^2} \Big[ \frac{i \pi \epsilon^2}{8 (\gamma_0 + \gamma_a)^2} - \frac{\pi \epsilon}{16 (\gamma_0 + \gamma_a)} \cr
  & & + \frac{i \gamma_0 (\gamma_0^2 + \gamma_a^2) \epsilon}{4 (\gamma_0^2 - \gamma_a^2)^2} + \frac{i \pi (2 \gamma_0^2 - \gamma_0 \gamma_a + \gamma_a^2)}{16 (\gamma_0 + \gamma_a)^2} - \frac{1}{8} + O(\frac{\gamma}{\epsilon}) \Big] \\
  \cr
  \eta^{aa}(\epsilon) & = & \int \frac{d p}{2 \pi} v_F p^2 \Bigg[ \frac{2(- v_F p + i \gamma_a) (v_F p + i \gamma_a) }{ \Omega(p)^2 + \kappa(p)^2} \cr
 & & + \frac{4 (v _F p \Omega(p) - \gamma_a \kappa(p)) (\Omega(p) + i \kappa(p)) (v_F p + i \gamma_a) }{ (\Omega(p)^2 + \kappa(p)^2)^2 }  \Bigg] \frac{ (\Omega(p) - i \kappa(p)) (v_F p - i \gamma_a) }{ \Omega(p)^2 + \kappa(p)^2} \cr
 & = & \frac{1}{2 \pi v_F^2} \Big[ \frac{i \pi \epsilon^2}{8 (\gamma_0 + \gamma_a)^2} - \frac{\pi \epsilon }{16 (\gamma_0 + \gamma_a)} \cr
 & & + \frac{i \gamma_0 (\gamma_0^2 + \gamma_a^2) \epsilon}{4 (\gamma_0^2 - \gamma_a^2)^2} - \frac{i \pi (\gamma_0 -3 \gamma_a) \gamma_a}{16 (\gamma_0 + \gamma_a)^2}  - \frac{\gamma_0^4 + 6 \gamma_0^2 \gamma_a^2 + \gamma_a^4}{8 (\gamma_0^2 - \gamma_a^2)^2} + O(\frac{\gamma}{\epsilon}) \Big] \\
 \cr
\eta^{aa}(\epsilon) - \eta^{00}(\epsilon) & = & \frac{1}{2 \pi v_F^2} \Big[ - \frac{\gamma_0^2 \gamma_a^2}{(\gamma_0^2 - \gamma_a^2)^2} - i \frac{\pi (\gamma_0 - \gamma_a)}{8 (\gamma_0 + \gamma_a)} \Big] \\
\cr
\eta^{0a}(\epsilon) & = & \int \frac{d p}{2 \pi} v_F p^2 \Bigg[ \frac{2(- v_F p + i \gamma_a) (\epsilon - i \gamma_0) }{ \Omega(p)^2 + \kappa(p)^2} \cr
 & & + \frac{4 (v _F p \Omega(p) - \gamma_a \kappa(p)) (\Omega(p) + i \kappa(p)) (\epsilon - i \gamma_0) }{ (\Omega(p)^2 + \kappa(p)^2)^2 }  \Bigg] \frac{ (\Omega(p) - i \kappa(p)) (v_F p - i \gamma_a) }{ \Omega(p)^2 + \kappa(p)^2} \cr
 & = & \frac{1}{2 \pi v_F^2} \Big[ i \frac{\pi  \epsilon^2}{8 (\gamma_0 + \gamma_a)^2} + \frac{\pi  \epsilon}{16 ( \gamma_0 +  \gamma_a)} + i \frac{\gamma_a \epsilon  (\gamma_0^2 + \gamma_a^2)}{4 ( \gamma_0^2 - \gamma_a^2)^2} \cr
 & & - i \frac{\pi (\gamma_0^2 -\gamma_0 \gamma_a + 2 \gamma_a^2) }{16 \left( \gamma_0 + \gamma_a \right)^2} + \frac{ \gamma_0^3 \gamma_a }{2 (\gamma_0^2 - \gamma_a^2)^2} + O(\frac{\gamma}{\epsilon})  \Big] \\
 \cr
 \eta^{a0}(\epsilon) & = & \int \frac{d p}{2 \pi} v_F p^2 \Bigg[ \frac{2(- v_F p + i \gamma_a) (v_F p + i \gamma_a) }{ \Omega(p)^2 + \kappa(p)^2} \cr
 & & + \frac{4 (v _F p \Omega(p) - \gamma_a \kappa(p)) (\Omega(p) + i \kappa(p)) (v_F p + i \gamma_a) }{ (\Omega(p)^2 + \kappa(p)^2)^2 }  \Bigg] \frac{ (\Omega(p) - i \kappa(p)) (\epsilon + i \gamma_0) }{ \Omega(p)^2 + \kappa(p)^2} \cr
 & = & \frac{1}{2 \pi v_F^2} \Big[ i \frac{\pi  \epsilon^2}{8 (\gamma_0 + \gamma_a)^2} + \frac{\gamma_0 \gamma_a \epsilon^2}{2 ( \gamma_0^2 -  \gamma_a^2)^2}  - \frac{3 \pi \epsilon}{16 ( \gamma_0 + \gamma_a)} + \frac{i \gamma_a (5 \gamma_0^2 + \gamma_a^2) \epsilon}{4 (\gamma_0^2 - \gamma_a^2)^2} \cr
 & & - \frac{i \pi \gamma_0 (\gamma_0 + 5 \gamma_a) }{16 \left( \gamma_0 + \gamma_a \right)^2} - \frac{3 \gamma_0^3 \gamma_a  + 2 \gamma_0 \gamma_a^3)}{2 (\gamma_0^2 - \gamma_a^2)^2} + O(\frac{\gamma}{\epsilon}) \Big] \\
 \cr
\eta^{0a}(\epsilon) - \eta^{a0}(\epsilon) & = & \frac{1}{2 \pi v_F^2} \Big[ \frac{\pi \epsilon}{4 (\gamma_0 + \gamma_a)} - \frac{i \gamma_0^2 \gamma_a \epsilon}{(\gamma_0^2 - \gamma_a^2)^2} + \frac{\gamma_0 \gamma_a (2 \gamma_0^2 + \gamma_a^2)}{(\gamma_0^2 - \gamma_a^2)^2} + O(\frac{\gamma}{\epsilon}) \Big]
\ea
\end{subequations}

Note that $\epsilon = \omega + \mu$ is the energy measured from $E_F$; hence, the DC conductivities will depend on $\eta^{ij}(E_F)$. We now re-write the SHE correlation function as a sum of several terms, $\pi^{z,(2)}_{xy}(\vec{k}, \omega) = \pi^{z,(2a)}_{xy}(\vec{k}, \omega)  + \pi^{z,(2b)}_{xy}(\vec{k}, \omega)$, where $\pi^{z,(2a)}_{xy}(\vec{k}, \omega)$ and $\pi^{z,(2b)}_{xy}(\vec{k}, \omega)$ are the $k_y \sigma^x$ and $k_x \sigma^y$ terms respectively. It is then necessary to Taylor expand $\Gbar^{(R)}(\vec{p} + \vec{k}, \epsilon) = \Gbar^{(R)}(\vec{p}, \epsilon) + k_i \tfrac{d \Gbar^{(R)}(\vec{p}, \epsilon)}{d p_i}$, and $\pi^{z, (2a1)}(\vec{k}, \omega)$ is the zeroth-order term, while $\pi^{z, (2a2)}(\vec{k}, \omega)$ and $\pi^{z, (2a3)}(\vec{k}, \omega)$ are the $k_x \tfrac{d \Gbar^{(R)}(\vec{p}, \epsilon)}{d p_x}$ and $k_y \tfrac{d \Gbar^{(R)}(\vec{p}, \epsilon)}{d p_y}$ terms respectively; thus, giving  $\pi^{z,(2a)}_{xy}(\vec{k}, \omega) = \pi^{z,(2a1)}_{xy}(\vec{k} = 0, \omega) + \pi^{z,(2a2)}_{xy}(\vec{k} = 0, \omega) + \pi^{z,(2a3)}_{xy}(\vec{k} = 0, \omega)$ and $\pi^{z,(2b)}_{xy}(\vec{k}, \omega) = \pi^{z,(2b1)}_{xy}(\vec{k} = 0, \omega) + \pi^{z,(2b2)}_{xy}(\vec{k} = 0, \omega) + \pi^{z,(2b3)}_{xy}(\vec{k} = 0, \omega)$. Finally, we make use of the chain rule $\tfrac{d \Gbar^{(R)}(\vec{p}, \epsilon)}{d p_i} = \tfrac{d \Gbar^{(R)}(\vec{p}, \epsilon)}{d p} \tfrac{\partial p}{ \partial p_i} + \tfrac{d \Gbar^{(R)}(\vec{p}, \epsilon)}{\partial \theta} \tfrac{\partial \theta}{ \partial p_i}$, which give $\pi^{z,(2a1)}_{xy}(\vec{k} = 0, \omega) = \pi^{z,(2a1P1)}_{xy}(\vec{k}, \omega) + \pi^{z,(2a1P2)}_{xy}(\vec{k}, \omega)$ respectively, with $\pi^{z,(2a1P1)}_{xy}(\vec{k}, \omega)$ and $\pi^{z,(2a1P2)}_{xy}(\vec{k}, \omega)$ being proportional to the $\tfrac{d \Gbar^{(R)}(\vec{p}, \epsilon)}{d p} \tfrac{\partial p}{ \partial p_i}$ and $\tfrac{d \Gbar^{(R)}(\vec{p}, \epsilon)}{\partial \theta} \tfrac{\partial \theta}{ \partial p_i}$ terms respectively. A similar procedure is carried out for the other terms, and we have symmetrized the expressions for $\pi^{z, (2a)}_{xy}(\vec{k}, \omega)$ and $\pi^{z, (2b)}_{xy}(\vec{k}, \omega)$ by doing a shift of variable $p_y + \tfrac{k_y}{2} \rightarrow p_y$ and $p_x + \tfrac{k_x}{2} \rightarrow p_x$ respectively. The results are shown below.
\begin{subequations}
\ba
\pi^{z,(2a)}_{xy}(\vec{k}, \omega) & = & \underset{\vec{k} \to 0}{\lim} \frac{2 i e v_F^2}{c} \int \frac{d \epsilon}{2 \pi i} \sum_{\vec{p}} \bl n_F(\epsilon) - n_F(\epsilon + \omega) \br 
\frac{p_y}{k_x} \cr
 & & \times \Tr \Big[ \Gbar^{(A)}(\vec{p} - \frac{\vec{k}}{2}, \epsilon) \, \sigma^x \, \Gbar^{R)}(\vec{p} + \frac{\vec{k}}{2}, \epsilon) \, \Gammabar^{(y)}(\vec{p}, \epsilon) \Big] \cr
& = & \pi^{z,(2a1)}_{xy}(\vec{k} = 0, \omega) + \pi^{z,(2a2)}_{xy}(\vec{k} = 0, \omega) + \pi^{z,(2a3)}_{xy}(\vec{k} = 0, \omega) \\
\cr
\pi^{z,(2a1)}_{xy}(\vec{k} = 0, \omega) & = &  \underset{\vec{k} \to 0}{\lim} \frac{2 i e v_F^2}{c} \frac{1}{k_x} \int \frac{d \epsilon}{2 \pi i} \sum_{\vec{p}} \bl n_F(\epsilon) - n_F(\epsilon + \omega) \br \cr
 & & \times \Tr \Big[ \Gbar^{(A)}(\vec{p}, \epsilon) \sigma^x \Gbar^{R)}(\vec{p}, \epsilon) \Gammabar^{(y)}(\vec{p}, \epsilon) \Big] p \sin \theta \cr
& = & 0 \\
\cr
\pi^{z,(2a2)}_{xy}(\vec{k} = 0, \omega) & = &  \underset{\vec{k} \to 0}{\lim} \frac{2 i e v_F^2}{c} \frac{k_x}{k_x} \int \frac{d \epsilon}{2 \pi i} \sum_{\vec{p}} \bl n_F(\epsilon) - n_F(\epsilon + \omega) \br \frac{p \sin \theta}{2} \cr
 & & \times \Bigg( \Tr \Big[ \Gbar^{(A)}(\vec{p}, \epsilon) \sigma^x \frac{\partial \Gbar^{(R)}(\vec{p}, \epsilon)}{\partial p_x} \Gammabar^{(y)}(\vec{p}, \epsilon) \Big] - \Tr \Big[ \frac{\partial \Gbar^{(A)}(\vec{p}, \epsilon)}{\partial p_x} \sigma^x \Gbar^{(R)}(\vec{p}, \epsilon) \Gammabar^{(y)}(\vec{p}, \epsilon) \Big] \Bigg) \cr
 & = & \underset{\vec{k} \to 0}{\lim} \frac{2 i e v_F^2}{c} \frac{k_x}{k_x} \int \frac{d \epsilon}{2 \pi i} \sum_{\vec{p}} \bl n_F(\epsilon) - n_F(\epsilon + \omega) \br \frac{p \sin \theta}{2} \cr
 & & \times \Bigg( \Tr \Big[ \Gbar^{(A)}(\vec{p}, \epsilon) \sigma^x \bl \frac{\partial \Gbar^{(R)}(\vec{p}, \epsilon)}{\partial p} \frac{\partial p}{\partial p_x} + \frac{\partial \Gbar^{(R)}(\vec{p}, \epsilon)}{\partial \theta} \frac{\partial \theta}{\partial p_x} \br \Gammabar^{(y)}(\vec{p}, \epsilon) \Big] \cr
 & & - \Tr \Big[ \bl \frac{\partial \Gbar^{(A)}(\vec{p}, \epsilon)}{\partial p} \frac{\partial p}{\partial p_x} + \frac{\partial \Gbar^{(A)}(\vec{p}, \epsilon)}{\partial \theta} \frac{\partial \theta}{\partial p_x} \br\sigma^x \Gbar^{(R)}(\vec{p}, \epsilon) \Gammabar^{(y)}(\vec{p}, \epsilon) \Big] \Bigg) \cr
& = & \pi^{z,(2a2P1)}_{xy}(\vec{k} = 0, \omega) + \pi^{z,(2a2P2)}_{xy}(\vec{k} = 0, \omega) \\
\cr
\pi^{z,(2a2P1)}_{xy}(\vec{k} = 0, \omega) & = &  \underset{\vec{k} \to 0}{\lim} \frac{2 i e v_F^2}{c} \frac{k_x}{k_x} \int \frac{d \epsilon}{2 \pi i} \sum_{\vec{p}} \bl n_F(\epsilon) - n_F(\epsilon + \omega) \br \frac{p \sin \theta}{2} \cr
 & & \times \Bigg( \Tr \Big[ \Gbar^{(A)}(\vec{p}, \epsilon) \sigma^x \bl \frac{\partial \Gbar^{(R)}(\vec{p}, \epsilon)}{\partial p} \frac{\partial p}{\partial p_x} \br \Gammabar^{(y)}(\vec{p}, \epsilon) \Big] \cr
 & & - \Tr \Big[ \bl \frac{\partial \Gbar^{(A)}(\vec{p}, \epsilon)}{\partial p} \frac{\partial p}{\partial p_x} \br \sigma^x \Gbar^{(R)}(\vec{p}, \epsilon) \Gammabar^{(y)}(\vec{p}, \epsilon) \Big] \Bigg) \cr
 & = &  \underset{\vec{k} \to 0}{\lim} \frac{2 i e v_F^2}{c} \frac{k_x}{k_x} \int \frac{d \epsilon}{2 \pi i} \sum_{\vec{p}} \bl n_F(\epsilon) - n_F(\epsilon + \omega) \br \frac{p \sin \theta}{2} \frac{\partial p}{\partial p_x} \cr
 & & \times \Bigg( \Tr \Big[ \Gbar^{(A)}(\vec{p}, \epsilon) \sigma^x \frac{\partial \Gbar^{(R)}(\vec{p}, \epsilon)}{\partial p} \Gammabar^{(y)}(\vec{p}, \epsilon) \Big] - \Tr \Big[ \frac{\partial \Gbar^{(A)}(\vec{p}, \epsilon)}{\partial p} \sigma^x \Gbar^{(R)}(\vec{p}, \epsilon) \Gammabar^{(y)}(\vec{p}, \epsilon) \Big] \Bigg) \cr
& = & \frac{2 i e v_F^2}{c} \int \frac{d \epsilon}{2 \pi i} \bl n_F(\epsilon) - n_F(\epsilon + \omega) \br \frac{1}{2} \cr
& & \times \frac{1}{4} \Big[ \Gamma^2_s(\epsilon) \bl 2 \eta^{aa}(\epsilon) - 2 (\eta^{aa}(\epsilon))^* \br \cr
& & + \Gamma^{0}_{p_y}(\epsilon) \bl \eta^{a0}(\epsilon) + \eta^{0a}(\epsilon) - (\eta^{a0}(\epsilon))^* - (\eta^{0a}(\epsilon))^* \br  \cr
& & + i \Gamma^{3}_{p_x}(\epsilon) \bl \eta^{a0}(\epsilon) - \eta^{0a}(\epsilon) + (\eta^{a0}(\epsilon))^* - (\eta^{0a}(\epsilon))^*  \br \Big] + O(\frac{\gamma}{E_F}) \\
\cr
\pi^{z,(2a2P2)}_{xy}(\vec{k} = 0, \omega) & = &  \underset{\vec{p} \to 0}{\lim} \frac{2 i e v_F^2}{c} \frac{k_x}{k_x} \int \frac{d \epsilon}{2 \pi i} \sum_{\vec{p}}  \bl n_F(\epsilon) - n_F(\epsilon + \omega) \br \frac{p \sin \theta}{2} \frac{\partial \theta}{\partial p_x} \cr
 & & \times \Bigg( \Tr \Big[ \Gbar^{(A)}(\vec{p}, \epsilon) \sigma^x \frac{\partial \Gbar^{(R)}(\vec{p}, \epsilon)}{\partial \theta} \Gammabar^{(y)}(\vec{p}, \epsilon) \Big] - \Tr \Big[ \frac{\partial \Gbar^{(A)}(\vec{p}, \epsilon)}{\partial \theta} \sigma^x \Gbar^{(R)}(\vec{p}, \epsilon) \Gammabar^{(y)}(\vec{p}, \epsilon) \Big] \Bigg) \cr
 & = & \frac{2 i e v_F^2}{c} \int \frac{d \epsilon}{2 \pi i}   \bl n_F(\epsilon) - n_F(\epsilon + \omega) \br \frac{1}{2} \cr
 & & \times \frac{1}{4} \Big[ 3 \Gamma^0_{p_y}(\epsilon) \bl \xi^{0a}(\epsilon) + \xi^{a0}(\epsilon) + i \xi^{3b}(\epsilon) + i \xi^{b3}(\epsilon) \br \cr
 & & + \Gamma^3_{p_x}(\epsilon) \bl  i \xi^{0a}(\epsilon) + i \xi^{a0}(\epsilon) \br \Big] + O(\frac{\gamma}{E_F}) \\
 \cr
 \pi^{z,(2a3)}_{xy}(\vec{k} = 0, \omega) & = & \underset{\vec{k} \to 0}{\lim} \frac{2 i e v_F^2}{c} \frac{k_y}{k_x} \int \frac{d \epsilon}{2 \pi i} \sum_{\vec{p}} \bl n_F(\epsilon) - n_F(\epsilon + \omega) \br \frac{p \sin \theta}{2} \cr
 & & \times \Bigg( \Tr \Big[ \Gbar^{(A)}(\vec{p}, \epsilon) \sigma^x \frac{\partial \Gbar^{(R)}(\vec{p}, \epsilon)}{\partial p_y} \Gammabar^{(y)}(\vec{p}, \epsilon) \Big] - \Tr \Big[ \frac{\partial \Gbar^{(A)}(\vec{p}, \epsilon)}{\partial p_y} \sigma^x \Gbar^{(R)}(\vec{p}, \epsilon) \Gammabar^{(y)}(\vec{p}, \epsilon) \Big] \Bigg) \cr
 & = & \underset{\vec{k} \to 0}{\lim} \frac{2 i e v_F^2}{c} \frac{k_y}{k_x} \int \frac{d \epsilon}{2 \pi i} \sum_{\vec{p}} \bl n_F(\epsilon) - n_F(\epsilon + \omega) \br \frac{p \sin \theta}{2} \cr
 & & \times \Bigg( \Tr \Big[ \Gbar^{(A)}(\vec{p}, \epsilon) \sigma^x \bl \frac{\partial \Gbar^{(R)}(\vec{p}, \epsilon)}{\partial p} \frac{\partial p}{\partial p_y} + \frac{\partial \Gbar^{(R)}(\vec{p}, \epsilon)}{\partial \theta} \frac{\partial \theta}{\partial p_y} \br \Gammabar^{(y)}(\vec{p}, \epsilon) \Big] \cr
 & & - \Tr \Big[ \bl \frac{\partial \Gbar^{(A)}(\vec{p}, \epsilon)}{\partial p} \frac{\partial p}{\partial p_y} + \frac{\partial \Gbar^{(A)}(\vec{p}, \epsilon)}{\partial \theta} \frac{\partial \theta}{\partial p_y} \br\sigma^x \Gbar^{(R)}(\vec{p}, \epsilon) \Gammabar^{(y)}(\vec{p}, \epsilon) \Big] \Bigg) \cr
& = & \pi^{z,(2a3P1)}_{xy}(\vec{k} = 0, \omega) + \pi^{z,(2a3P2)}_{xy}(\vec{k} = 0, \omega) \\
\cr
\pi^{z,(2a3P1)}_{xy}(\vec{k} = 0, \omega) & = &  \underset{\vec{p} \to 0}{\lim} \frac{2 i e v_F^2}{c} \frac{k_y}{k_x} \int \frac{d \epsilon}{2 \pi i} \sum_{\vec{p}} \bl n_F(\epsilon) - n_F(\epsilon + \omega) \br \frac{p \sin \theta}{2} \frac{\partial p}{\partial p_y} \cr
 & & \times \Bigg( \Tr \Big[ \Gbar^{(A)}(\vec{p}, \epsilon) \sigma^x \frac{\partial \Gbar^{(R)}(\vec{p}, \epsilon)}{\partial p} \Gammabar^{(y)}(\vec{p}, \epsilon) \Big] - \Tr \Big[ \frac{\partial \Gbar^{(A)}(\vec{p}, \epsilon)}{\partial p} \sigma^x \Gbar^{(R)}(\vec{p}, \epsilon) \Gammabar^{(y)}(\vec{p}, \epsilon) \Big] \Bigg) \cr
& = & \frac{2 i e v_F^2}{c} \int \frac{d \epsilon}{2 \pi i}   \bl n_F(\epsilon) - n_F(\epsilon + \omega) \br \frac{1}{2} \cr
& & \times \frac{1}{4} \Big[ \Gamma^1_s(\epsilon) \bl 4 \eta^{00}(\epsilon) - 4 (\eta^{00}(\epsilon))^* - 2 \eta^{aa}(\epsilon) +  2 (\eta^{aa}(\epsilon))^* \br \cr
& & + \Gamma^{0}_{p_x}(\epsilon) \bl \eta^{a0}(\epsilon) + \eta^{0a}(\epsilon) - (\eta^{a0}(\epsilon))^* + (\eta^{0a}(\epsilon))^* \br  \cr
& & + \Gamma^{3}_{p_y}(\epsilon) \bl - 3 \eta^{a0}(\epsilon) + 3 \eta^{0a}(\epsilon) - 3 (\eta^{a0}(\epsilon))^* + 3 (\eta^{0a}(\epsilon))^*  \br \Big] + O(\frac{\gamma}{E_F}) \\
\cr
\pi^{z,(2a3P2)}_{xy}(\vec{k} = 0, \omega) & = &  \underset{\vec{p} \to 0}{\lim} \frac{2 i e v_F^2}{c} \frac{k_y}{k_x} \int \frac{d \epsilon}{2 \pi i} \sum_{\vec{p}} \bl n_F(\epsilon) - n_F(\epsilon + \omega) \br \frac{p \sin \theta}{2} \frac{\partial \theta}{\partial p_y} \cr
 & & \times \Bigg( \Tr \Big[ \Gbar^{(A)}(\vec{p}, \epsilon) \sigma^x \frac{\partial \Gbar^{(R)}(\vec{p}, \epsilon)}{\partial \theta} \Gammabar^{(y)}(\vec{p}, \epsilon) \Big] - \Tr \Big[ \frac{\partial \Gbar^{(A)}(\vec{p}, \epsilon)}{\partial \theta} \sigma^x \Gbar^{(R)}(\vec{p}, \epsilon) \Gammabar^{(y)}(\vec{p}, \epsilon) \Big] \Bigg) \cr
 & = & \frac{2 i e v_F^2}{c} \int \frac{d \epsilon}{2 \pi i}   \bl n_F(\epsilon) - n_F(\epsilon + \omega) \br \frac{1}{2} \cr
 & & \times \frac{1}{4} \Big[ \Gamma^0_{p_x}(\epsilon) \bl \xi^{0a}(\epsilon) - \xi^{a0}(\epsilon) \br - \Gamma^3_{p_y}(\epsilon) \bl  \xi^{0a}(\epsilon) + \xi^{a0}(\epsilon) \br \Big] + O(\frac{\gamma}{E_F}) \\
 \cr
\pi^{z,(2b)}_{xy}(\vec{k}, \omega) & = & \underset{\vec{k} \to 0}{\lim} -\frac{2 i e v_F^2}{c} \int \frac{d \epsilon}{2 \pi i} \sum_{\vec{p}} \bl n_F(\epsilon) - n_F(\epsilon + \omega) \br 
\frac{p_x}{k_x} \cr
 & & \times \Tr \Big[ \Gbar^{(A)}(\vec{p} - \frac{\vec{k}}{2}, \epsilon) \, \sigma^y \, \Gbar^{R)}(\vec{p} + \frac{\vec{k}}{2}, \epsilon) \, \Gammabar^{(y)}(\vec{p}, \epsilon) \Big] \cr
& = & \pi^{z,(2b1)}_{xy}(\vec{k} = 0, \omega) + \pi^{z,(2b2)}_{xy}(\vec{k} = 0, \omega) + \pi^{z,(2b3)}_{xy}(\vec{k} = 0, \omega) \\
\cr
\pi^{z,(2b1)}_{xy}(\vec{k} = 0, \omega) & = &  \underset{\vec{k} \to 0}{\lim} -\frac{2 i e v_F^2}{c} \frac{1}{k_x} \int \frac{d \epsilon}{2 \pi i} \sum_{\vec{p}}  \bl n_F(\epsilon) - n_F(\epsilon + \omega) \br \cr
 & & \times \Tr \Big[ \Gbar^{(A)}(\vec{p}, \epsilon) \sigma^y \Gbar^{R)}(\vec{p}, \epsilon) \Gammabar^{(y)}(\vec{p}, \epsilon) \Big] p \cos \theta \cr
& = & 0 \\
\cr
\pi^{z,(2b2)}_{xy}(\vec{k} = 0, \omega) & = &  \underset{\vec{k} \to 0}{\lim} -\frac{2 i e v_F^2}{c} \frac{k_x}{k_x} \int \frac{d \epsilon}{2 \pi i} \sum_{\vec{p}} \bl n_F(\epsilon) - n_F(\epsilon + \omega) \br \frac{p \cos \theta}{2} \cr
 & & \times \Bigg( \Tr \Big[ \Gbar^{(A)}(\vec{p}, \epsilon) \sigma^y \frac{\partial \Gbar^{(R)}(\vec{p}, \epsilon)}{\partial p_x} \Gammabar^{(y)}(\vec{p}, \epsilon) \Big] - \Tr \Big[ \frac{\partial \Gbar^{(A)}(\vec{p}, \epsilon)}{\partial p_x} \sigma^y \Gbar^{(R)}(\vec{p}, \epsilon) \Gammabar^{(y)}(\vec{p}, \epsilon) \Big] \Bigg) \cr
 & = & \underset{\vec{k} \to 0}{\lim} -\frac{2 i e v_F^2}{c} \frac{k_x}{k_x} \int \frac{d \epsilon}{2 \pi i} \sum_{\vec{p}} \bl n_F(\epsilon) - n_F(\epsilon + \omega) \br \frac{p \cos \theta}{2} \cr
 & & \times \Bigg( \Tr \Big[ \Gbar^{(A)}(\vec{p}, \epsilon) \sigma^y \bl \frac{\partial \Gbar^{(R)}(\vec{p}, \epsilon)}{\partial p} \frac{\partial p}{\partial p_x} + \frac{\partial \Gbar^{(R)}(\vec{p}, \epsilon)}{\partial \theta} \frac{\partial \theta}{\partial p_x} \br \Gammabar^{(y)}(\vec{p}, \epsilon) \Big] \cr
 & & - \Tr \Big[ \bl \frac{\partial \Gbar^{(A)}(\vec{p}, \epsilon)}{\partial p} \frac{\partial p}{\partial p_x} + \frac{\partial \Gbar^{(A)}(\vec{p}, \epsilon)}{\partial \theta} \frac{\partial \theta}{\partial p_x} \br\sigma^y \Gbar^{(R)}(\vec{p}, \epsilon) \Gammabar^{(y)}(\vec{p}, \epsilon) \Big] \Bigg) \cr
& = & \pi^{z,(2b2P1)}_{xy}(\vec{k} = 0, \omega) + \pi^{z,(2b2P2)}_{xy}(\vec{k} = 0, \omega) \\
\cr
\pi^{z,(2b2P1)}_{xy}(\vec{k} = 0, \omega) & = &  \underset{\vec{p} \to 0}{\lim} - \frac{2 i e v_F^2}{c} \frac{k_x}{k_x} \int \frac{d \epsilon}{2 \pi i} \sum_{\vec{p}}  \bl n_F(\epsilon) - n_F(\epsilon + \omega) \br \frac{p \cos \theta}{2} \frac{\partial p}{\partial p_x} \cr
 & & \times \Bigg( \Tr \Big[ \Gbar^{(A)}(\vec{p}, \epsilon) \sigma^y \frac{\partial \Gbar^{(R)}(\vec{p}, \epsilon)}{\partial p} \Gammabar^{(y)}(\vec{p}, \epsilon) \Big] - \Tr \Big[ \frac{\partial \Gbar^{(A)}(\vec{p}, \epsilon)}{\partial p} \sigma^y \Gbar^{(R)}(\vec{p}, \epsilon) \Gammabar^{(y)}(\vec{p}, \epsilon) \Big] \Bigg) \cr
& = & - \frac{2 i e v_F^2}{c} \int \frac{d \epsilon}{2 \pi i}   \bl n_F(\epsilon) - n_F(\epsilon + \omega) \br \frac{1}{2} \cr
& & \times \frac{1}{4} \Big[ \Gamma^2_s(\epsilon) \bl 4 \eta^{00}(\epsilon) - 4(\eta^{00}(\epsilon))^* - 2 \eta^{aa}(\epsilon) + 2 (\eta^{aa}(\epsilon))^* \br \cr
& & + \Gamma^{0}_{p_y}(\epsilon) \bl \eta^{a0}(\epsilon) + \eta^{0a}(\epsilon) - (\eta^{a0}(\epsilon))^* - (\eta^{0a}(\epsilon))^* \br  \cr
& & + i \Gamma^{3}_{p_x}(\epsilon) \bl 3\eta^{0a}(\epsilon) -3 \eta^{a0}(\epsilon) + 3 (\eta^{0a}(\epsilon))^* - 3 (\eta^{a0}(\epsilon))^*   \br \Big] + O(\frac{\gamma}{E_F}) \\
\cr
\pi^{z,(2b2P2)}_{xy}(\vec{k} = 0, \omega) & = &  \underset{\vec{p} \to 0}{\lim} - \frac{2 i e v_F^2}{c} \frac{k_x}{k_x} \int \frac{d \epsilon}{2 \pi i} \sum_{\vec{p}}  \bl n_F(\epsilon) - n_F(\epsilon + \omega) \br \frac{p \cos \theta}{2} \frac{\partial \theta}{\partial p_x} \cr
 & & \times \Bigg( \Tr \Big[ \Gbar^{(A)}(\vec{p}, \epsilon) \sigma^y \frac{\partial \Gbar^{(R)}(\vec{p}, \epsilon)}{\partial \theta} \Gammabar^{(y)}(\vec{p}, \epsilon) \Big] - \Tr \Big[ \frac{\partial \Gbar^{(A)}(\vec{p}, \epsilon)}{\partial \theta} \sigma^y \Gbar^{(R)}(\vec{p}, \epsilon) \Gammabar^{(y)}(\vec{p}, \epsilon) \Big] \Bigg) \cr
 & = & - \frac{2 i e v_F^2}{c} \int \frac{d \epsilon}{2 \pi i}   \bl n_F(\epsilon) - n_F(\epsilon + \omega) \br \frac{1}{2} \cr
 & & \times \frac{1}{4} \Big[ \Gamma^0_{p_y}(\epsilon) \bl \xi^{0a}(\epsilon) - \xi^{a0}(\epsilon) \br - \Gamma^3_{p_x}(\epsilon) \bl  i \xi^{0a}(\epsilon) + i \xi^{a0}(\epsilon) \br \Big] + O(\frac{\gamma}{E_F}) \\
 \cr
 \pi^{z,(2b3)}_{xy}(\vec{k} = 0, \omega) & = & \pi^{z,(2b3P1)}_{xy}(\vec{k} = 0, \omega) + \pi^{z,(2b3P2)}_{xy}(\vec{k} = 0, \omega) \\
\cr
\pi^{z,(2b3P1)}_{xy}(\vec{k} = 0, \omega) & = &  \underset{\vec{k} \to 0}{\lim} - \frac{2 i e v_F^2}{c} \frac{k_y}{k_x} \int \frac{d \epsilon}{2 \pi i} \sum_{\vec{p}}  \bl n_F(\epsilon) - n_F(\epsilon + \omega) \br \frac{p \cos \theta}{2} \frac{\partial p}{\partial p_y} \cr
 & & \times \Bigg( \Tr \Big[ \Gbar^{(A)}(\vec{p}, \epsilon) \sigma^y \frac{\partial \Gbar^{(R)}(\vec{p}, \epsilon)}{\partial p} \Gammabar^{(y)}(\vec{p}, \epsilon) \Big] - \Tr \Big[ \frac{\partial \Gbar^{(A)}(\vec{p}, \epsilon)}{\partial p} \sigma^y \Gbar^{(R)}(\vec{p}, \epsilon) \Gammabar^{(y)}(\vec{p}, \epsilon) \Big] \Bigg) \cr
& = & - \frac{2 i e v_F^2}{c} \int \frac{d \epsilon}{2 \pi i}   \bl n_F(\epsilon) - n_F(\epsilon + \omega) \br \frac{1}{2} \cr
& & \times \frac{1}{4} \Big[ \Gamma^1_s(\epsilon) \bl 2 \eta^{aa}(\epsilon) -  2 (\eta^{aa}(\epsilon))^* \br \cr
& & + \Gamma^{0}_{p_x}(\epsilon) \bl \eta^{a0}(\epsilon) + \eta^{0a}(\epsilon) - (\eta^{a0}(\epsilon))^* - (\eta^{0a}(\epsilon))^* \br  \cr
& & + \Gamma^{3}_{p_y}(\epsilon) \bl \eta^{a0}(\epsilon) - \eta^{0a}(\epsilon) + (\eta^{a0}(\epsilon))^*- (\eta^{0a}(\epsilon))^*  \br \Big] + O(\frac{\gamma}{E_F}) \\
\cr
\pi^{z,(2b3P2)}_{xy}(\vec{k} = 0, \omega) & = &  \underset{\vec{k} \to 0}{\lim} - \frac{2 i e v_F^2}{c} \frac{k_y}{k_x} \int \frac{d \epsilon}{2 \pi i} \sum_{\vec{p}}  \bl n_F(\epsilon) - n_F(\epsilon + \omega) \br \frac{p \sin \theta}{2} \frac{\partial \theta}{\partial p_y} \cr
 & & \times \Bigg( \Tr \Big[ \Gbar^{(A)}(\vec{p}, \epsilon) \sigma^x \frac{\partial \Gbar^{(R)}(\vec{p}, \epsilon)}{\partial \theta} \Gammabar^{(y)}(\vec{p}, \epsilon) \Big] - \Tr \Big[ \frac{\partial \Gbar^{(A)}(\vec{p}, \epsilon)}{\partial \theta} \sigma^x \Gbar^{(R)}(\vec{p}, \epsilon) \Gammabar^{(y)}(\vec{p}, \epsilon) \Big] \Bigg) \cr
 & = & - \frac{2 i e v_F^2}{c} \int \frac{d \epsilon}{2 \pi i}   \bl n_F(\epsilon) - n_F(\epsilon + \omega) \br \frac{1}{2} \cr
 & & \times \frac{1}{4} \Big[ 3 \Gamma^0_{p_x}(\epsilon) \bl \xi^{a0}(\epsilon) - \xi^{0a}(\epsilon) \br - \Gamma^3_{p_y}(\epsilon) \bl  \xi^{0a}(\epsilon) + \xi^{a0}(\epsilon) \br \Big] + O(\frac{\gamma}{E_F})
\ea
\end{subequations}

Therefore, summing up all the different contributions, we finally obtain the SHE correlation function,
\ba
\label{eqn: Piz2xy result}
\pi^{z, (2)}(\vec{p} =0, \omega) & = & \frac{2 i e v_F^2}{c} \int \frac{d \epsilon}{2 \pi i} (n_F(\epsilon) - n_F(\epsilon + \omega)) \cr
 & & \times \frac{1}{2} \Bigg[ \Gamma^0_{p_x}(\epsilon) \Big( \xi^{0a}(\epsilon) - \xi^{a0}(\epsilon) \Big) + \Gamma^0_{p_y}(\epsilon) \Big( \xi^{a0}(\epsilon) - \xi^{0a}(\epsilon) \Big) \cr
 & & \Gamma^1_s(\epsilon) \Big( \eta^{00}(\epsilon) - \eta^{aa}(\epsilon) - (\eta^{00}(\epsilon))^* + (\eta^{aa}(\epsilon))^* \Big) \cr
 & & + \Gamma^2_s(\epsilon) \Big( \eta^{aa}(\epsilon) - \eta^{00}(\epsilon) - (\eta^{aa}(\epsilon))^* + (\eta^{00}(\epsilon))^*  \Big) \cr
 & & + \Gamma^3_{p_x}(\epsilon) \Big( i \eta^{a0}(\epsilon) - i \eta^{0a}(\epsilon) + i (\eta^{a0}(\epsilon))^* - i (\eta^{0a}(\epsilon))^* \Big) \cr
 & & + \Gamma^3_{p_y}(\epsilon) \Big( \eta^{0a}(\epsilon) - \eta^{a0}(\epsilon) + (\eta^{0a}(\epsilon))^* - (\eta^{a0}(\epsilon))^* \Big) + O(\frac{\gamma}{\epsilon}) \Bigg] 
\ea

Using the results for $\xi^{ij}(\omega)$ and $\eta^{ij}(\omega)$ from above, where $\xi^{0a}(\omega) - \xi^{a0}(\omega) = - \tfrac{i \pi}{2 \pi v_F^2}$, $Im[ \eta^{aa}(\omega)- \eta^{00}(\omega)] = - \tfrac{1}{2 \pi v_F^2} \tfrac{\pi (\gamma_0 - \gamma_a)}{8 (\gamma_0 + \gamma_a)}$, and $Re[\eta^{0a}(\omega) - \eta^{a0}(\omega)] = \tfrac{1}{2 \pi v_F^2} \tfrac{\pi \omega}{4 (\gamma_0 + \gamma_a)} = \tfrac{\pi N_0(\omega)}{4 (\gamma_0 + \gamma_a)}$, we see that the main $O(\tfrac{1}{\gamma})$ contributions come from the $\Gamma^3_{p_x}(\omega)$ scattering channel.

The uniform DC longitudinal charge and spin-Hall conductivity are given by $\sigma_{yy} = - \Lim{\omega \rightarrow 0} \Lim{\vec{k} \rightarrow 0} Im \left[ \frac{\pi_{yy}(\vec{k}, \omega)}{\omega} \right]$, $\sigma^{z}_{xy} = -\Lim{\omega \rightarrow 0} \Lim{\vec{k} \rightarrow 0} Im \left[ \frac{\pi^{z}_{xy}(\vec{k}, \omega)}{\omega} \right]$, and keeping only the $O(\tfrac{1}{\gamma})$ terms, they are,
\ba
\label{eqn: spin charge conductivities}
\sigma_{yy} & = & \frac{1}{2 \pi} \bl e v_F \br^2 Re \Big[ 2 \Gamma^2_0(E_F) \, \xi^{00}(E_F) \Big] \cr
 & = & \bl e v_F \br^2 \frac{N_0(E_F)}{2 \gamma_t} + O \bl \frac{\gamma}{E_F} \br \\
\sigma^{z, (2)}_{xy} & = &  \frac{\hbar e v_F^2}{\pi} Im \Big[ i \Gamma^3_{p_x}(E_F) \big[ Re[ \eta^{0a}(E_F) - \eta^{a0}(E_F)] \big] \Big] \cr
 & = & - \hbar e v_F^2 \frac{N_0(E_F)}{2 \gamma_t} \frac{\gamma_{s}}{\gamma_0 + \gamma_a} + O \bl \frac{\gamma}{E_F} \br  \\
\sigma^y_y & = & \frac{\hbar e v_F}{2 \pi} Re \Big[ 2 \Gamma^2_s(E_F) \xi^{00}(E_F) \Big]\cr
 & = & \hbar e v_F \frac{N_0(E_F)}{2 \gamma_t} + O \bl \frac{\gamma}{E_F} \br 
\ea

Hence, we see that the SHE is driven by scattering between the $s$ and $p$-wave electrons due to the symmetric spin-flip $T^S$ term, which occurs at  $3^{rd}$-order in perturbation. Eq.~\ref{eqn: Gamma3px}, $\Gamma^3_{p_x}(E_F) =  - \frac{\gamma_{s}}{\gamma_t} - i \frac{\gamma_{31} + \gamma_{asym,2}}{\gamma_t} + \frac{\gamma_{3s} \gamma_{asym,1}}{2 \gamma_t (\gamma_0 + \gamma_a)} $, shows that the asymmetric spin-flip term $T^A$ also contributes but as a sub-leading term, .

\bibliography{SHE_bibliography}